\documentclass[11pt]{article}
\pdfoutput=1
\usepackage[T1]{fontenc}
\usepackage[utf8]{inputenc}
\usepackage{graphicx}
\usepackage{caption}
\usepackage{subcaption}
\usepackage{bbold}
\usepackage{array}
\usepackage[margin=1in]{geometry}
\usepackage{amsmath,amsthm,amssymb}
\usepackage{indentfirst}
\usepackage{algorithm}
\usepackage[noend]{algpseudocode}
\usepackage{tikz,tkz-tab}
\usepackage[toc,page]{appendix}
\usepackage{setspace}
\begin{document}
\title{Instantaneous order impact and high-frequency strategy optimization in limit order books}
\author{Federico Gonzalez\thanks{Corresponding author: fgonzale@andrew.cmu.edu}{}~ and Mark Schervish, \\
Department of Statistics, \\
Carnegie Mellon University, Pittsburgh, PA}
\date{}
\maketitle

\begin{abstract}
\noindent
We propose a limit order book model with dynamics that account for both the impact of the most recent order and volume imbalance. To model these effects jointly we introduce a discrete Markov chain model. We then find the policy for optimal order choice and control. The optimal policy derived uses limit orders, cancellations and market orders. It looks to avoid non-execution and adverse selection risk simultaneously.  Using ultra high-frequency data from the NASDAQ stock exchange we compare our policy with other submission strategies that use a subset of all available order types and show that ours significantly outperforms them. 
\vspace{2mm}
\\
\noindent
\emph{Keywords}: market microstructure, limit order books, markov decision processes, adverse selection, non-execution risk 
\end{abstract}

\section{Introduction}\label{intro}
In most modern financial markets, trading activity is organized by a limit order book (LOB) structure. Market participants can interact with the LOB through three types of orders: market orders, limit orders and cancellations. A market order (MO) is an order to immediately buy or sell a specific volume of shares at the best price available. A limit order (LO) specifies not only the volume to buy or sell, but also the worst acceptable price, these orders wait in the LOB to be matched by a counterpart for a trade. Cancellations (CO) of pending limit orders can occur at any time. Most LOBs match orders by a price-time priority rule. This means that pending LOs are matched to MOs first based on their price, and then on their submission time. Understanding and modelling the evolution of LOBs is of major practical importance and an extensive literature in the topic exists. See Gould, et al. (2013) and Abergel et al. (2016) for a thorough survey of modelling techniques for LOBs.
\par 
This paper focuses on two problems. First we extend existing LOB models to incorporate the influence in order flow of both recent book events and the shape of the LOB. Then, under these improved dynamics we show how to solve the problem of optimal acquisition of one lot of shares using all order types dynamically depending on the LOB conditions.
\par
The strategic behavior of market participants is always evolving. However, many statistical properties have been observed in the order flow they produce on LOBs. The intensity rate of the arrival of all order types is known to depend on the volume at the best bid and ask prices, this was captured by the model proposed by Huang et al. (2015) with state-dependent Poisson order flows. However, the time durations between events are not independent and display non-trivial correlation patterns. Rambaldi et al. (2016) and Bacry et al. (2016) successfully model these complex interactions between the arrivals of different order types with multivariate Hawkes processes. 
\par 
We perform an empirical analysis of high-frequency data showing that the distribution of the type of the next order depends on the volume at the best prices as well as the type of the most recent order. Our first contribution is the introduction of a discrete Markov chain model for the LOB dynamics that incorporates simultaneously both of these effects. We find that different orders impact immediate future order flow in significantly different ways. In our analysis, we describe in detail this ``instantaneous impact'', which we interpret as the immediate strategic reaction of market participants to each new order arrival. 
\par
The framework we propose is related to some of the existing queuing models for LOBs. However, there are several differences relative to some of the most well known models in the area: Cont et al. (2010) and Huang (2015). We focus on a simplified version which models only the best bid and ask and assumes constant spread of size one. There is a significant group of financial assets where this assumption is realistic, the so called ``large-tick'' assets. See for example Dayri and Rosenbaum (2015). The volume levels at the best prices along with the type of the last order observed, represent the state space of our Markov chain. We allow the transition probabilities between states to depend on the volumes at the best prices as well as the type of the last order observed. This allows our model to incorporate the influence of the LOB configuration and the historical flows. The dynamics of our model occur in discrete steps according to ``event time,'' i.e. the discrete clock advances with the arrival of each order. This allows us to estimate only the parameters of interest to understand the evolution of the LOB, while capturing the most useful properties of the order flow.
\par
Widespread availability of high quality market data has led to an explosive growth in the literature of high-frequency algorithmic trading. Modern markets are extremely competitive, and market participants try to exploit all available information in order to trade optimally. One well known source of predictive information about the future mid-price is the volume imbalance at the best quotes. For example Gould and Bonart (2016) perform a large scale empirical analysis in this topic. Studies show how to incorporate this predictive power into high-frequency trading strategies. See Lehalle and Mounjid (2016), Donelly and Gan (2017), Jaquier and Liu (2017). The decay of the predictive power of imbalance is not well understood. Cartea, et al. (2015c) use imbalance to predict the mid-price after a fixed time horizon and Lehalle and Mounjid (2016) predict the mid-price a fixed number of orders in the future.  However, our empirical analysis in the Appendix shows that imbalance is useful to predict the next two mid-price changes, but then it quickly loses its power as a predictor for further mid-price changes. It is interesting to note that this fact is implied by most queuing models, which assume that both queues are renewed randomly after a mid-price change.  
\par
Our primary focus is on the question of how to place orders in a LOB optimally under the dynamics described by our model. We frame the problem of optimal acquisition of one share as a Markov decision process. The optimal strategy derived involves all order types MOs, LOs and COs. The acquisition price is benchmarked against the next mid-price after a change. This choice is in part justified by the duration study of the imbalance signal and it shortens the optimization horizons significantly so that the optimal strategy can be numerically computed very efficiently.  Lehalle and Mounjid (2016) use a different benchmark in their solution to the problem of optimally controlling a limit order.  But both benchmarks cause the optimal strategies to cancel limit orders that face a high risk of adverse selection, i.e. buying (selling) when the price is about to go down (up). Since our strategy also incorporates MOs, it aggressively takes liquidity by submitting market buy (sell) orders when the non-execution risk of a limit buy (sell) order placed in the book is high and the price is about to move adversely up (down). 
\par 
The rest of the paper is organized as follows. In Section \ref{data}, we describe in detail the data used in our empirical study. In Section \ref{analysis1},  we presents some findings of LOB order flow and introduce our Markov chain model for the LOB dynamics.  In Section \ref{placement}, we introduce a Markov decision process framework for optimal execution and an algorithm to derive the optimal strategy. In Section \ref{strategy}, we present the derived optimal strategy calibrated to real market data and discuss some economic insights gained. We summarize our results in Section \ref{conclusion}.  In Appendix \ref{imbalance}, we present a detailed analysis of the decay of the predictive power of imbalance.
\section{Data}\label{data}
We use data from the NASDAQ Historical TotalView-ITCH database from January 2nd to March 31st of 2015. This dataset includes all MO, LOsand CO arrivals timestamped up to nanosecond precision. On the NASDAQ exchange, each stock is traded on a separate LOB. The smallest permissible price interval between different orders, also known as tick size, is equal to \$0.01. Although this minimal price interval is fixed, the prices of different stocks vary widely. A key differentiator in market activity for each particular stock is the ratio of the tick size and its price. Liquid stocks where this ratio is large are usually known as large-tick stocks, since the tick size is large relative to the stock price. Many statistical properties of this group of stocks have been described in the LOB literature. The most relevant ones for our study are that the spread, i.e. the difference between the bid and ask prices, is almost always equal to one tick and that most of the order submission activity occurs at the bid and ask price levels. See Dayri and Rosenbaum (2015) for a detailed analysis of large-tick stocks.
\par
The empirical analysis and results of model fitting are based on data from Microsoft (MSFT) and Intel (INTC), however we have verified that our main conclusions hold for other large-tick stocks. Our choice is based on sorting stocks by volume traded. From this sorted list we selected the top stocks with a spread almost always constant and equal to \$0.01.
\par
As is common in the analysis of high-frequency LOB data, we exclude market activity from the first 30 minutes after market opening and the last 30 minutes before market close, as well as any activity outside market hours. This is done to avoid the impact on our study of abnormal noisy trading behavior that occurs during those periods. The volumes for all order types (MO, LO and CO) are normalized by a factor equal to the median of the volume of all order types. We only track the volume at the bid and ask prices. Therefore, all LOs and COs at other price levels are ignored. We set an upper bound for the normalized volume at both price levels in order for the state space of the LOB to be finite.
\section{Order flow analysis and modelling}\label{analysis1}
Empirical studies of high-frequency LOB data have shown that there are at least two major factors in determining the submission rates of all order types, the volume levels at the bid and ask prices, see Huang et al. (2015), as well as historical order flows, see Rambaldi et al. (2016). Our goal in this section is to study how these two effects interact and incorporate them in a simple discrete Markov chain model for LOB dynamics.
\par
Let $(V^{b}_{t}, V^{a}_{t})$ denote the normalized volume levels at the bid and ask prices in the LOB. Also define volume imbalance denoted by $I_{t}$ as
\begin{equation}
I_{t} = \frac{V^{b}_{t} - V^{a}_{t}}{V^{b}_{t} + V^{a}_{t}}
\end{equation}
\par
It is well known that $I_{t}$ influences future price dynamics. Gould and Bonart (2016) show in a large sample study that $I_{t}$ is a strong predictor of the next mid-price change for large-tick stocks. This naturally implies that $I_{t}$ impacts the order submission strategies of market participants. Huang et al. (2015) study in detail how the submission rates of all order types depend on the volume at each price level. 
\par
Numerous studies in the financial literature have documented the complex interactions between the arrival rates of different order types. The most remarkable and universal characteristic is the self-exciting nature of these processes. Several models that account for these features exist. Bacry et al. (2016) propose a multivariate Hawkes process to model all events at the first level of the LOB. 
\par
It is clear then that the intensities of all order types depend both on recent orders and on the state of the LOB. Here, we present an analysis of both of these effects jointly. Let $e_{t}$ denote the type of the last order to modify the normalized volume at the bid or ask queue at time $t$. The six order types we will consider are market buy (MB), market sell (MS), limit buy (LB), limit sell (LS), cancel buy (CB) and cancel sell (CS) respectively. We will take the tuple $(V^{b}_{t}, V^{a}_{t}, e_{t})$ as a summary of both the state of the LOB and the historical order flow. Although simple, we will show that this representation allows us to capture most of the statistical relationships of interest. Finally let $t_{i}, i = 1, 2, \ldots$ denote the clock times of the arrival of all orders that modify the volume at the best quotes. 
\par
Assume for now that the spread of the LOB is constant and equal to one tick. For a given summary tuple $(V^{b}_{t}, V^{a}_{t}, e_{t})$, $V^{b}_{t}$ and $V^{a}_{t}$ can only increase or decrease with the arrivals of an order as in (\ref{arrow}). By our discussion above, the probability that the next order observed is of a given type should depend on the both the volume levels and the last order. 
\begin{equation}\label{arrow}
(V^{b}_{t_{i}}, V^{a}_{t_{i}}, e_{t_{i}}) \xrightarrow{e_{t_{i+1}}} (V^{b}_{t_{i+1}}, V^{a}_{t_{i+1}}, e_{t_{i+1}})
\end{equation}
\par
To verify this hypothesis we first discretize the range of $I_{t} \in [ -1, 1]$ into 5 equal size bins. Let $D_{t}$ be this discretized version of imbalance. We wish to estimate the probability of observing each of the six order types binned by what we will refer to as the reduced state of the LOB $(D_{t}, e_{t})$. Equivalently, for each state $(V_{b}^{t}, V_{a}^{t}, e_{t})$ we reduce it into $(D_{t}, e_{t})$ and estimate the probability of each order type that will first modify it. There are 30 such reduced pairs (5 imbalance levels and 6 order types) and 5 probabilities per reduced state (6 order types but probabilities must add to 1). 
\par
\begin{figure}[h!]
    \centering
    \begin{subfigure}[t]{0.32\textwidth}
        \centering
        \includegraphics[width=\linewidth]{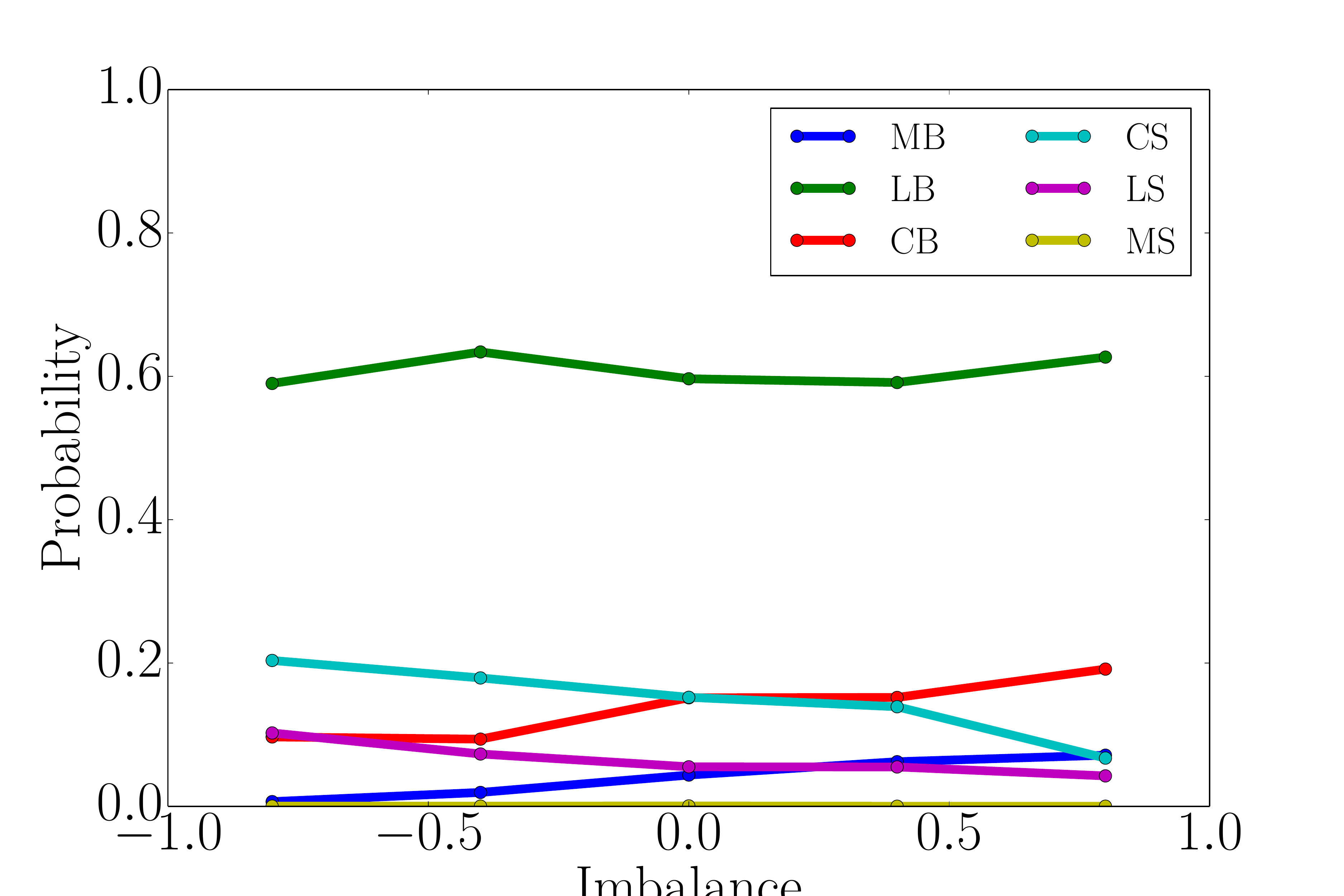} 
        \caption{Previous event $ = $ MB} \label{fig:timing1}
    \end{subfigure}
    \hfill
    \begin{subfigure}[t]{0.32\textwidth}
        \centering
        \includegraphics[width=\linewidth]{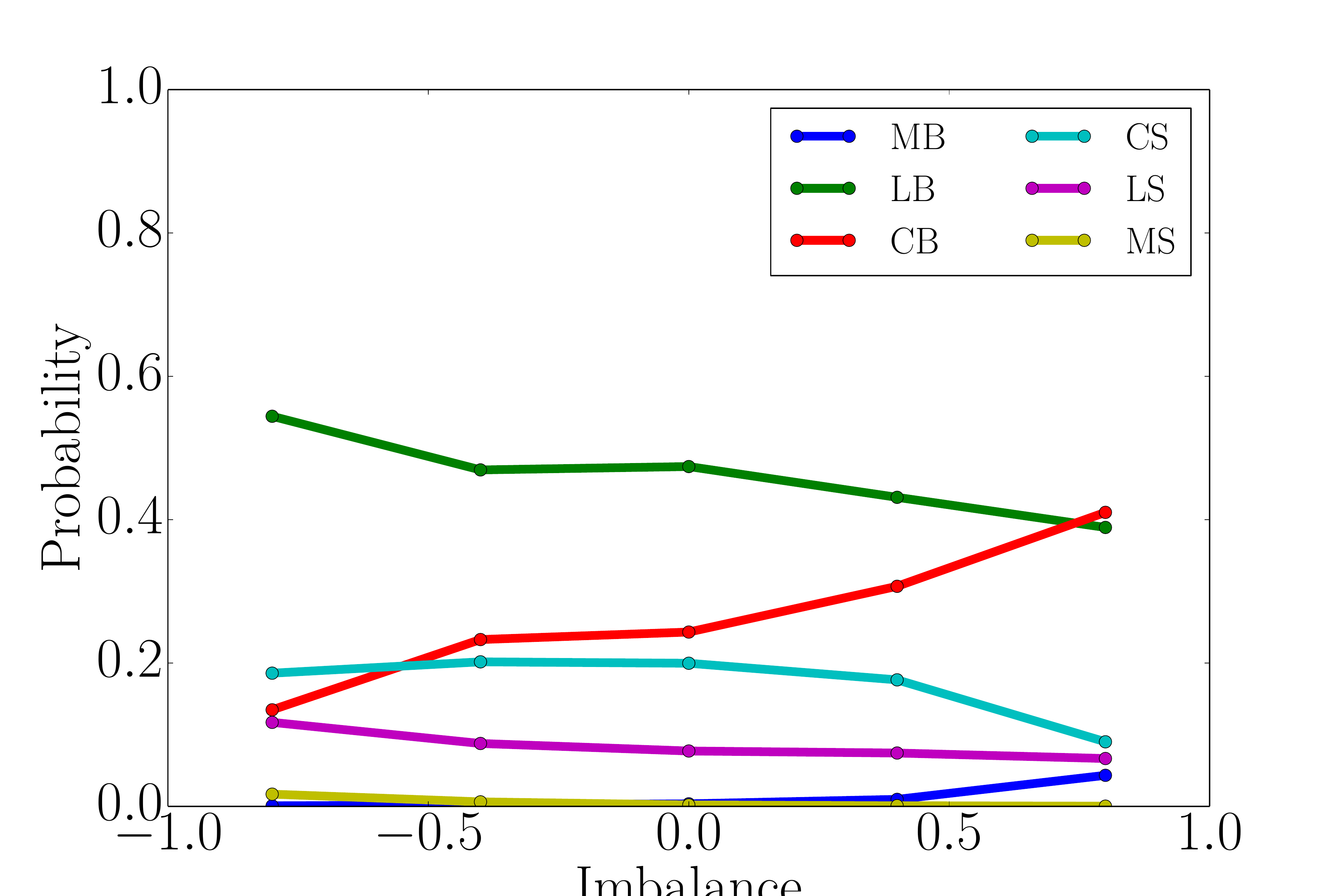} 
        \caption{Previous event $ = $ LB} \label{fig:timing2}
    \end{subfigure}
    \hfill
        \begin{subfigure}[t]{0.32\textwidth}
        \centering
        \includegraphics[width=\linewidth]{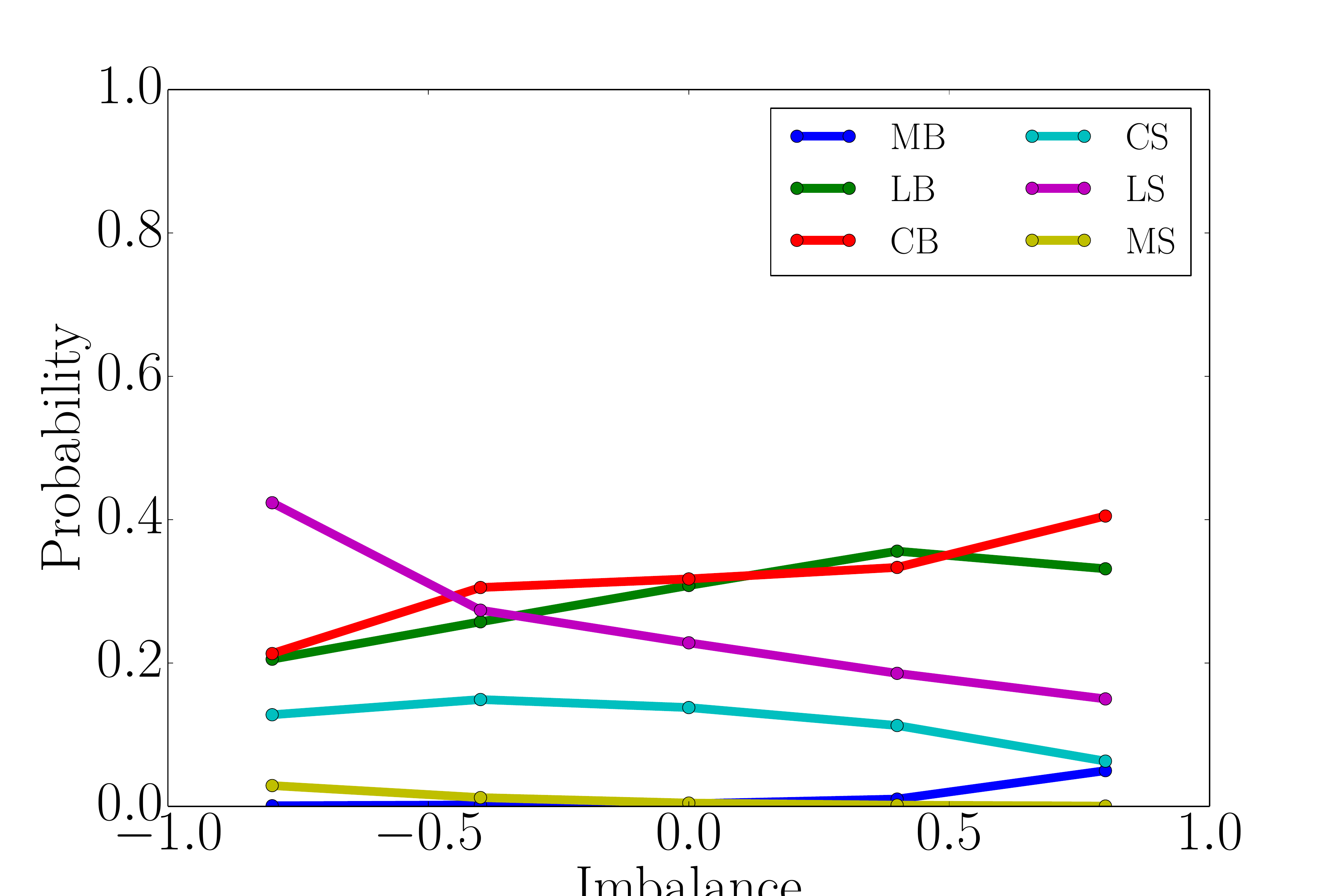} 
        \caption{Previous event $ = $ CB} \label{fig:timing3}
    \end{subfigure}
    \vspace{1cm}
    \begin{subfigure}[t]{0.32\textwidth}
        \centering
        \includegraphics[width=\linewidth]{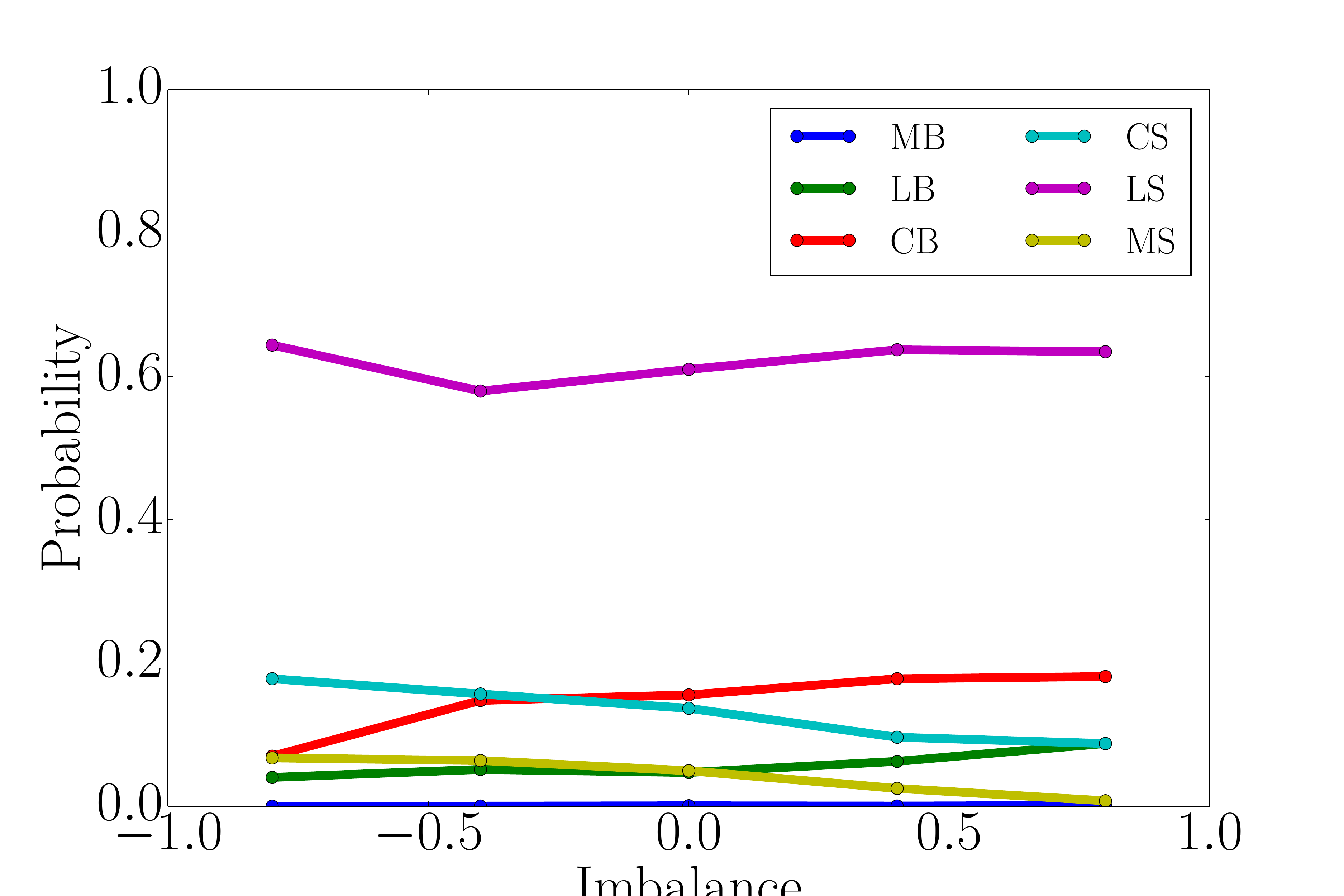} 
        \caption{Previous event $ = $ MS} \label{fig:timing4}
    \end{subfigure}
    \vspace{1cm}
        \hfill
    \begin{subfigure}[t]{0.32\textwidth}
        \centering
        \includegraphics[width=\linewidth]{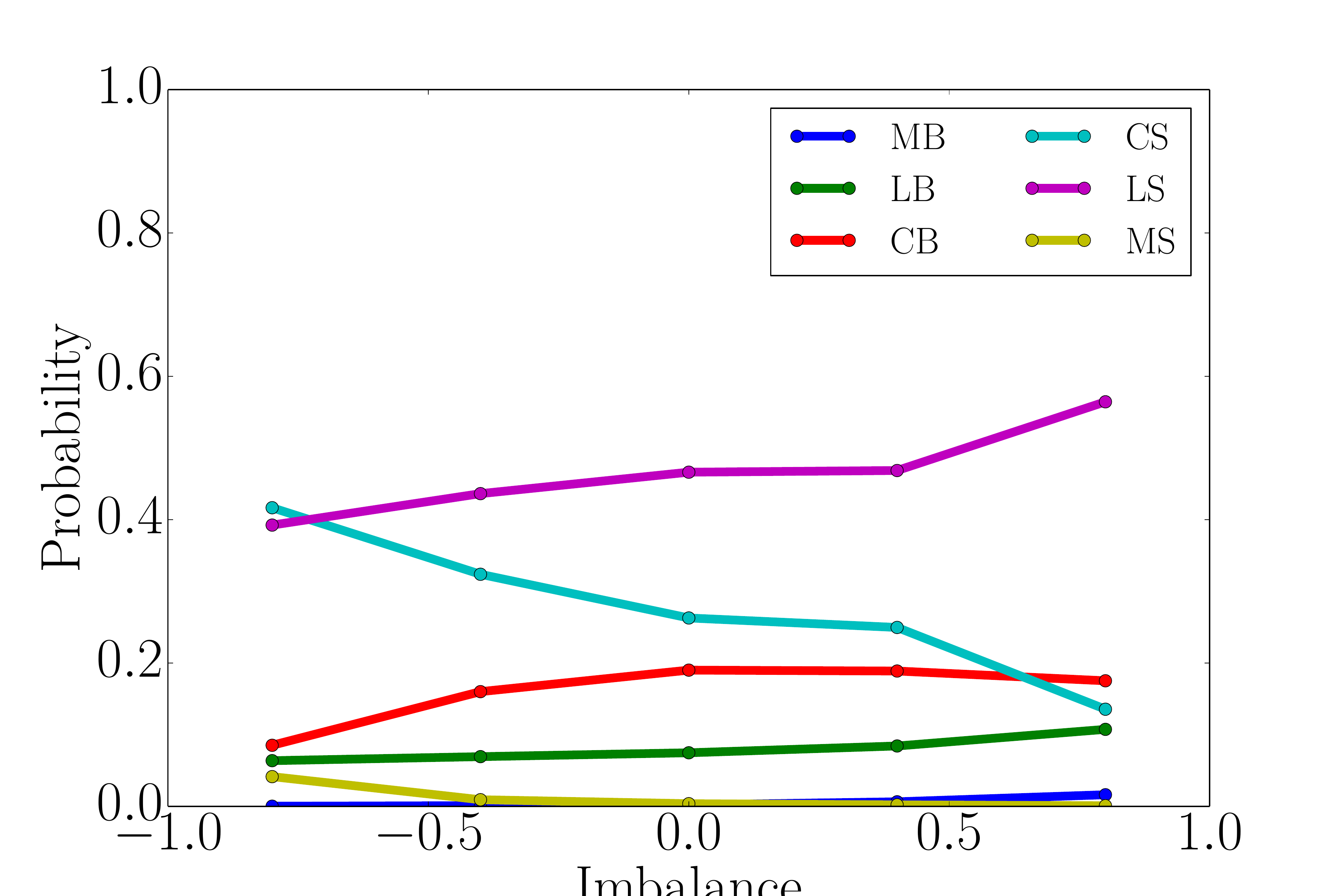} 
        \caption{Previous event $ = $ LS} \label{fig:timing5}
    \end{subfigure}
    	\hfill
    	    \begin{subfigure}[t]{0.32\textwidth}
        \centering
        \includegraphics[width=\linewidth]{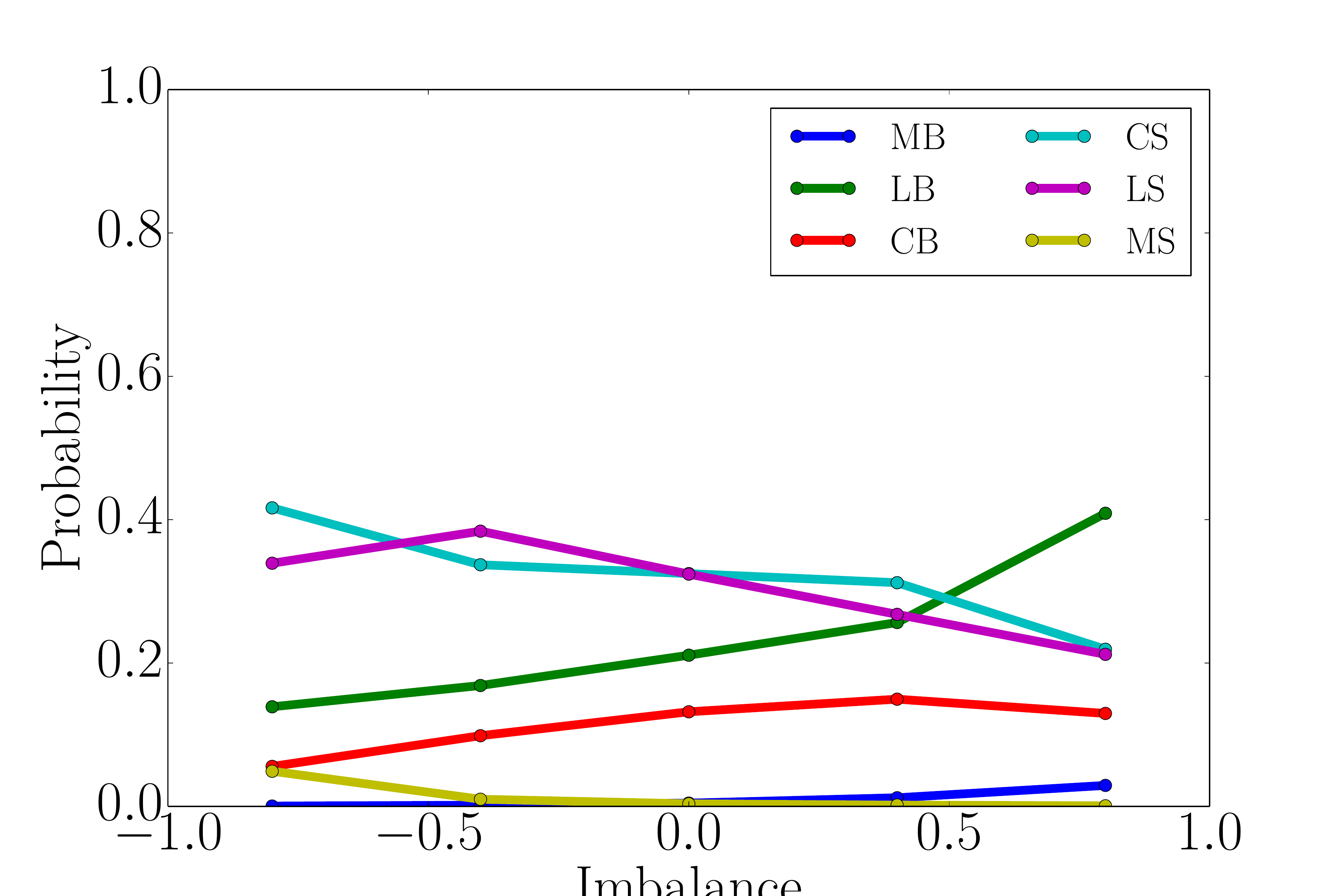} 
        \caption{Previous event $ = $ CS} \label{fig:timing6}
    \end{subfigure}
    \vspace{-4\baselineskip}
    \captionsetup{justification=centering}
    \caption{For MSFT data, each figure shows the empirical probability of each order type conditioned on the state of imbalance of the imbalance of the LOB and the last order observed \\ (a) MB, (b) LB, (c) CB, (d) CS, (e) LS and (f) MS}
\label{fig:transitions_MSFT}
\end{figure}
\begin{figure}[h!]
    \centering
    \begin{subfigure}[t]{0.32\textwidth}
        \centering
        \includegraphics[width=\linewidth]{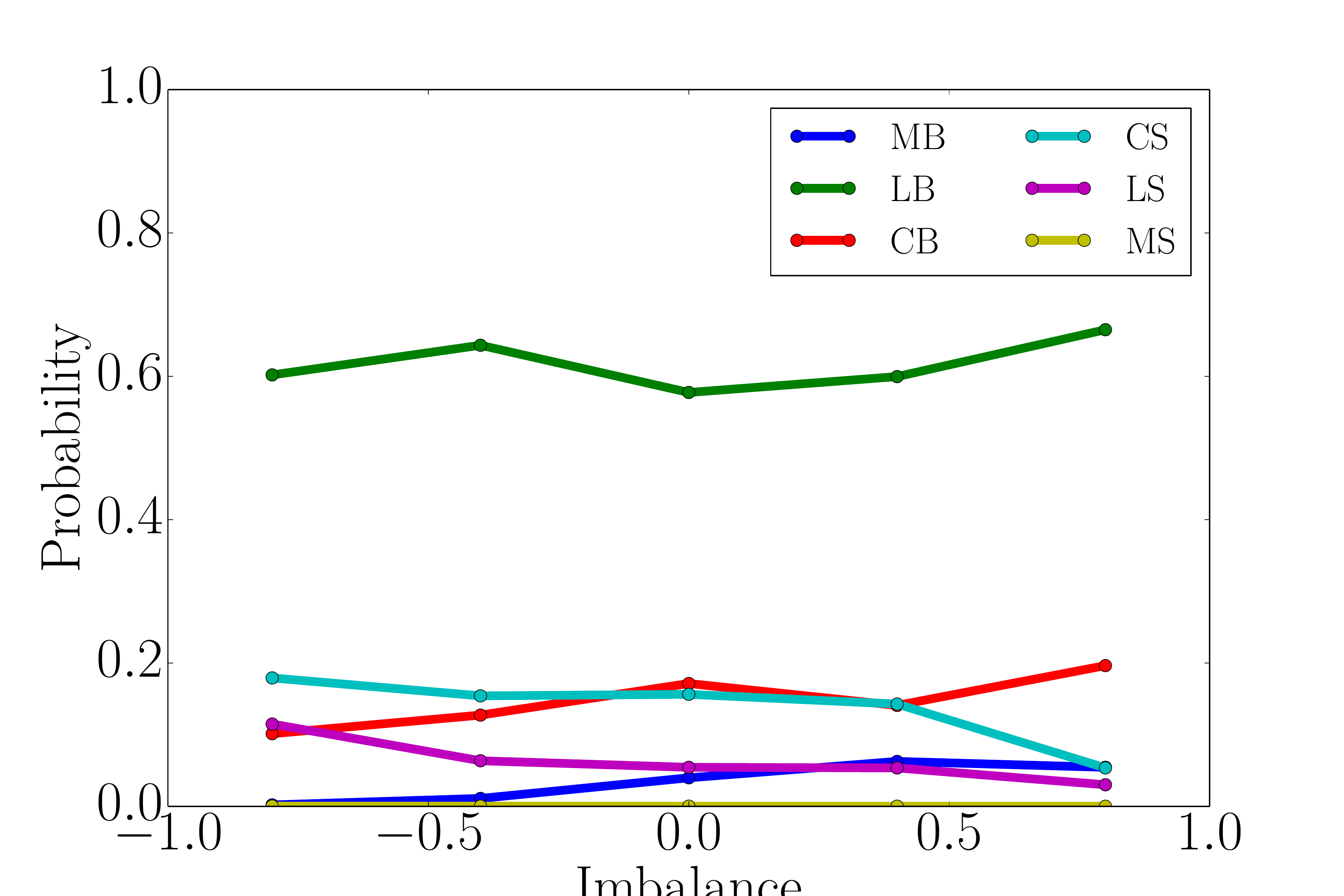} 
        \caption{Previous event $ = $ MB} \label{fig:timing7}
    \end{subfigure}
    \hfill
    \begin{subfigure}[t]{0.32\textwidth}
        \centering
        \includegraphics[width=\linewidth]{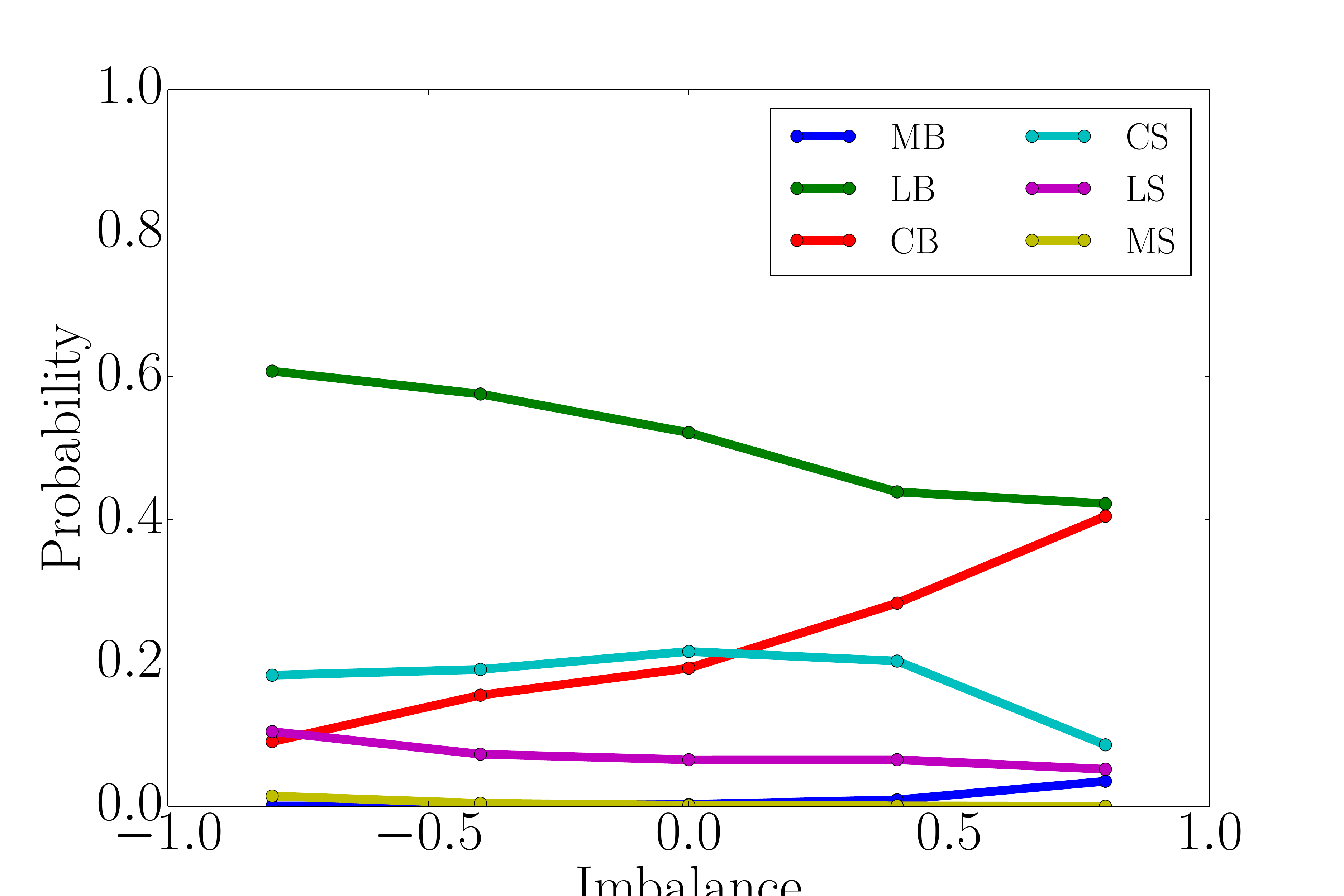} 
        \caption{Previous event $ = $ LB} \label{fig:timing8}
    \end{subfigure}
    \hfill
        \begin{subfigure}[t]{0.32\textwidth}
        \centering
        \includegraphics[width=\linewidth]{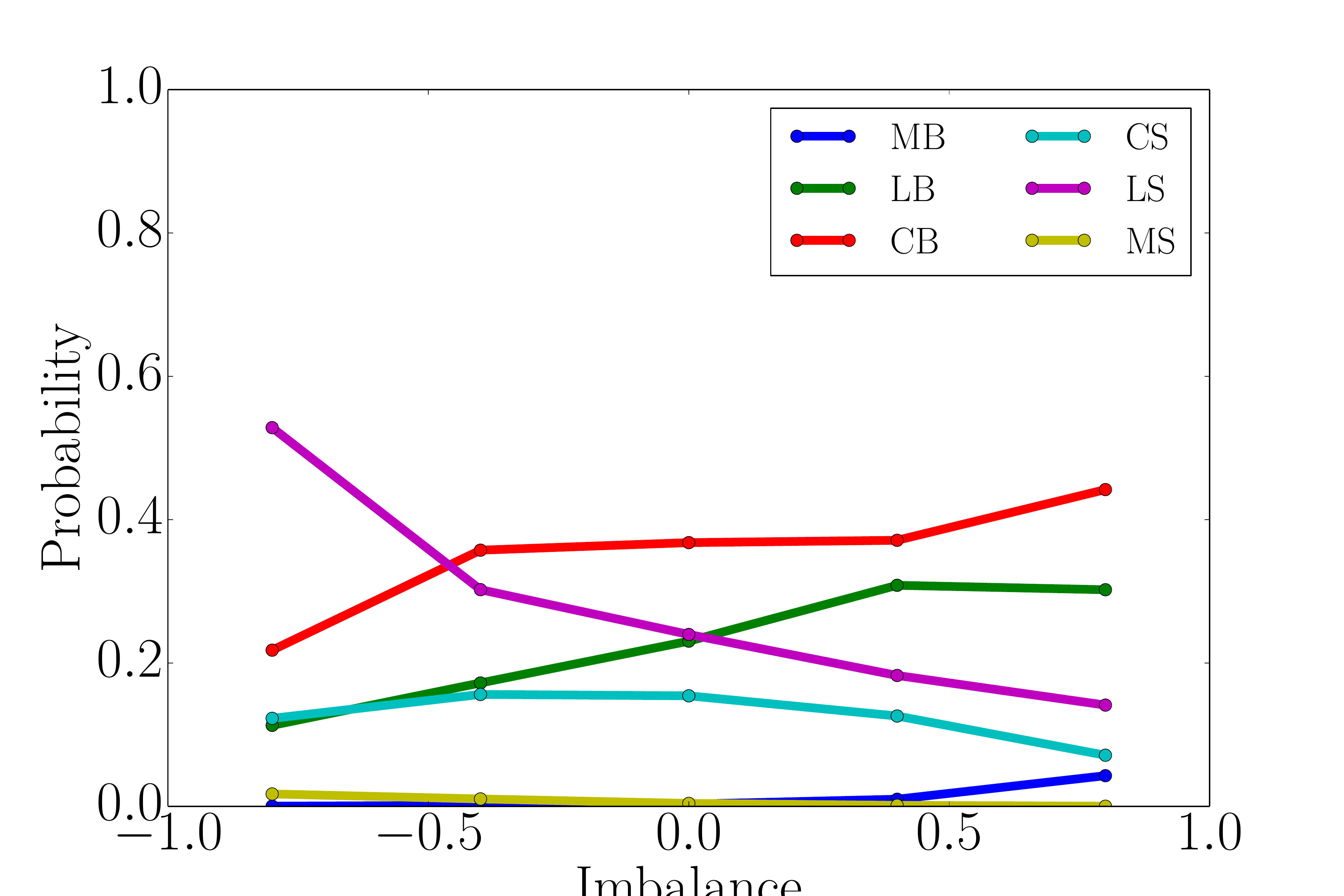} 
        \caption{Previous event $ = $ CB} \label{fig:timing9}
    \end{subfigure}
    \vspace{1cm}
    \begin{subfigure}[t]{0.32\textwidth}
        \centering
        \includegraphics[width=\linewidth]{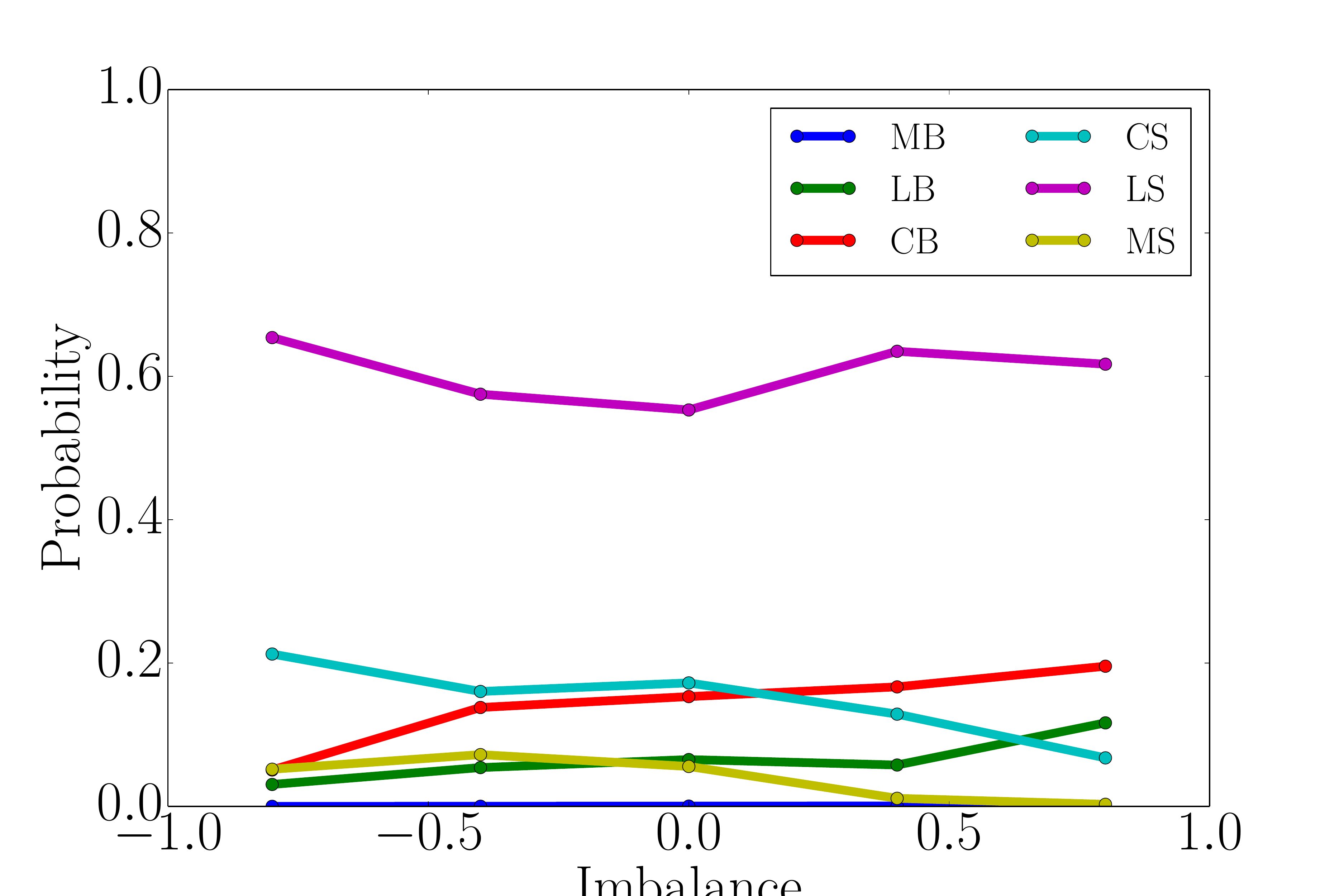} 
        \caption{Previous event $ = $ MS} \label{fig:timing10}
    \end{subfigure}
    \vspace{1cm}
        \hfill
    \begin{subfigure}[t]{0.32\textwidth}
        \centering
        \includegraphics[width=\linewidth]{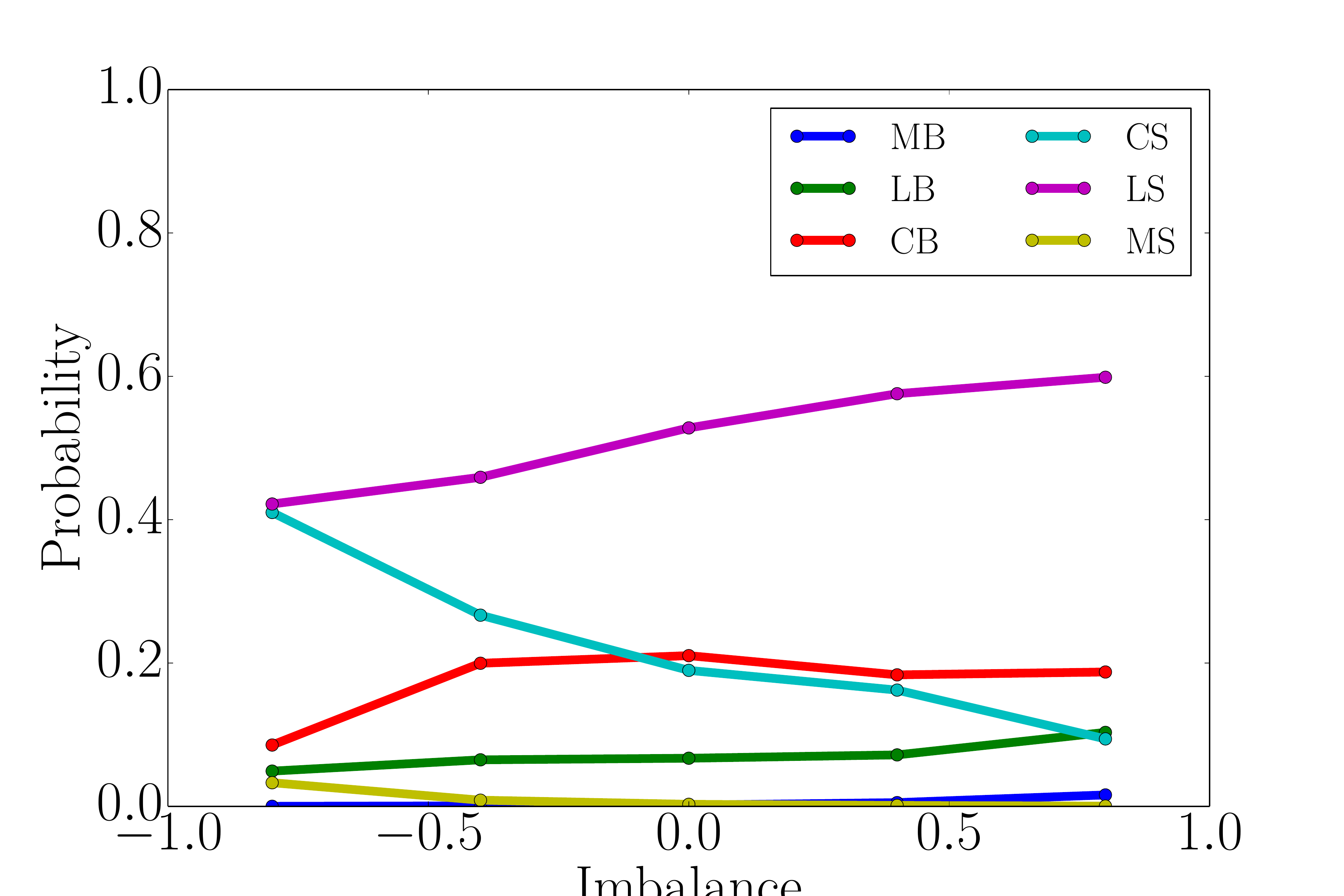} 
        \caption{Previous event $ = $ LS} \label{fig:timing11}
    \end{subfigure}
    	\hfill
    \begin{subfigure}[t]{0.32\textwidth}
        \centering
        \includegraphics[width=\linewidth]{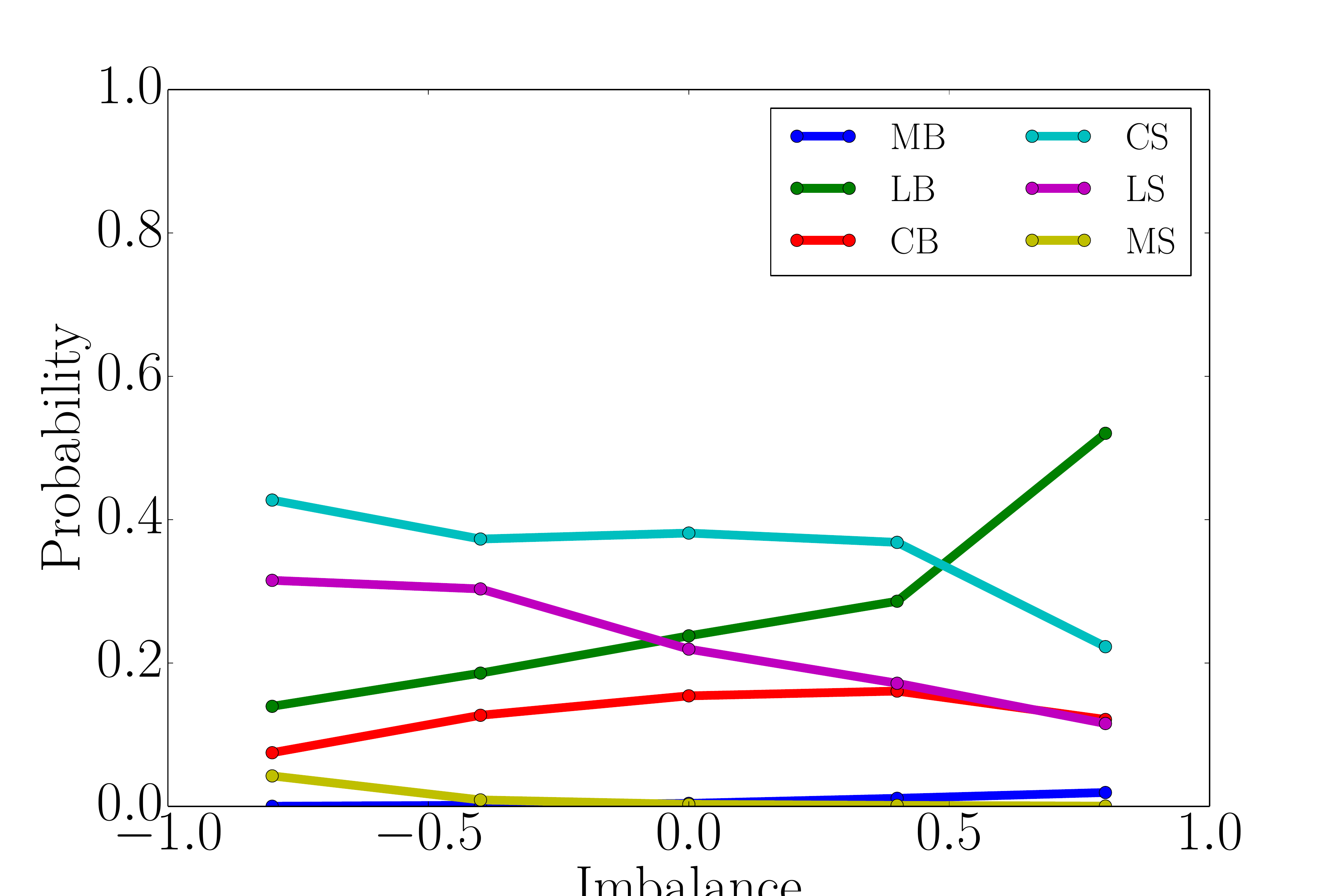} 
        \caption{Previous event $ = $ CS} \label{fig:timing12}
    \end{subfigure}
    \vspace{-3\baselineskip}
    \captionsetup{justification=centering}
    \caption{For INTC data, each figure shows the empirical probability of each order type conditioned on the state of imbalance of the imbalance of the LOB and the last order observed \\ (a) MB, (b) LB, (c) CB, (d) CS, (e) LS and (f) MS}
\label{fig:transitions_INTC}
\end{figure}

For a given reduced state $s_{t} = (D_{t}, e_{t})$ and last observed order type $h =$ MB, MS, LB, LS, CB and CS denote by $p_{s_{t}}^{h}$ the probability that a LOB with reduced state $s_{t}$ is first modified by an order of type $h$. The MLE of each of these probabilities is the empirical proportion of counts of observed order $h$ for a LOB in reduced state $s_t$. The results for the stocks in our sample are shown in Fig. \ref{fig:transitions_MSFT} and \ref{fig:transitions_INTC}. We split the reduced states by last order observed and plot the proportion of observed counts for each order type versus the discretized imbalance. For example in Fig. \ref{fig:transitions_MSFT} (a) given that the last order observed was a MB we plot 6 curves that show how the probability of next observing each order type varies with the discretized imbalance.
\par
In the following sections we give a qualitative discussion of the results and their implications as well as the statistical significance of our estimations.

\subsection{Statistical significance}\label{hypothesis}
\begin{figure}[h!]
    \begin{subfigure}[t]{0.45\textwidth}
        \centering
        \includegraphics[width=\linewidth]{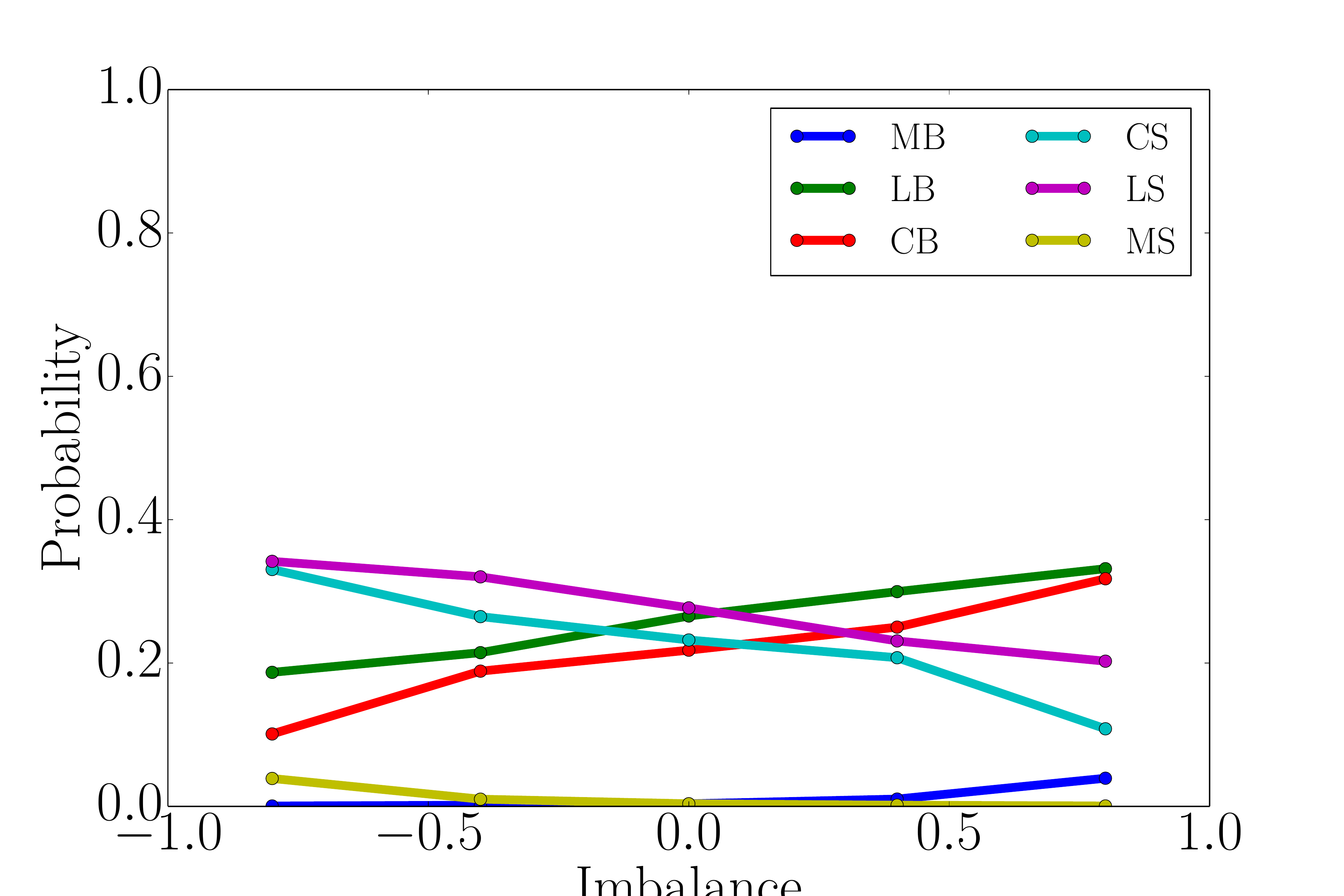} 
        \caption{MSFT} 
        \label{MSFT_no_prev}
    \end{subfigure}
    	\hfill
    \begin{subfigure}[t]{0.45\textwidth}
        \centering
        \includegraphics[width=\linewidth]{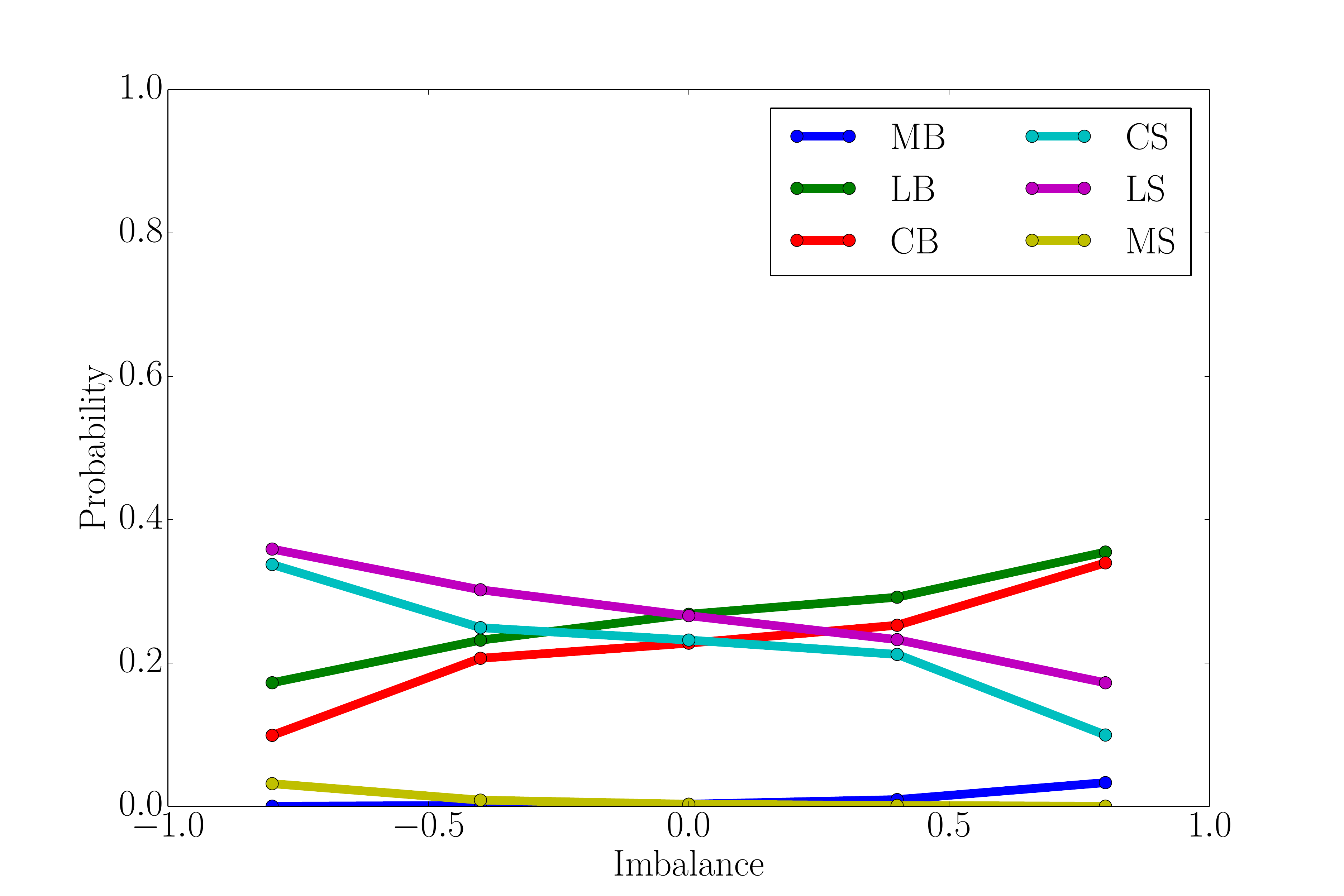} 
        \caption{INTC} 
        \label{INTC_no_prev}
    \end{subfigure}
    \captionsetup{justification=centering}
    \caption{Empirical probabilities conditioned on imbalance for data on MSFT (left) and INTC (right)}
    \label{unconditional}
\end{figure}

We would like to test whether the last order observed has an instantaneous effect on the order submission strategies of market participants. In Fig. \ref{unconditional} we show the empirical probabilities of observing each order type but only conditioning on discretized imbalance $D_{t}$. If the last order observed had no effect then panels (a)-(f) in Fig. \ref{fig:transitions_MSFT} and \ref{fig:transitions_INTC} should all be similar to Fig. \ref{unconditional} respectively. Informally it is clear from the figures that this effect souldn't be ignored. To validate this formally we proceed as follows. 
\par
For each reduced state $(d, e)$ and order type $f$, where $d \in D = \{1,2,3,4,5\}$ the set of discretized imbalance values and $e, f \in E = \{ \text{MB, LB, CB, CS, LS, MS} \}$ the set of order types, let $p^{f}_{(d,e)}$ denote the probability that the next observed order is of type $f$ conditioned on the LOB being in reduced state $(d,e)$. Similarly denote by $p^{f}_{d}$ the probability that the next observed order type is $f$ conditioned on the LOB having discretized imbalance $d$. We propose the following hypothesis test:
\begin{align}
H_{0} : p^{f}_{d} &= p^{f}_{(d,e)}, \forall d \in D, e,f \in E \\
H_{A} : p^{f}_{(d,e)} &= p^{f}_{(d,e')}, \exists d \in D, e, e', f \in E
\end{align}
\par 
By our specifications this amounts to testing two nested multinomial models. One with a set of parameters for each imbalance level and another with parameters for each combination of imbalance and previous order type. Let $\hat{p}^{f}_{(d,e)}$ and $\hat{p}^{f}_{d}$ by the MLE estimates for $p^{f}_{(d,e)}$ and $p^{f}_{d}$ defined before. 
Also let $N^{f}_{(d,e)}$ be the observed counts for order type $f$ being observed next when the LOB state is $(d,e)$. The generalized likelihood ratio test statistic is given by
\begin{equation}
-2\log\Lambda = 2 \sum_{d \in D} \sum_{e \in E} \sum_{f \in E} \left( N^{f}_{(d,e)} \log \frac{\hat{p}^{f}_{(d,e)}}{\hat{p}^{f}_{d}}\right)
\end{equation}
\par
Under the null hypothesis the distribution of $-2\log\Lambda$ is approximately $\chi^{2}_{125}$. The results of this test are shown in Table \ref{hypothesis_table}. It can be seen that there is significant evidence to reject the null on both stocks in our sample. We remark that the observed order counts are not uniformly distributed across reduced states. 
\begin{table}[h!]
\centering
\caption{Nested models hypothesis test}
\begin{tabular}{c | c | c | c}
\hline
Stock  & $-2\log\Lambda$ & p-value & \# Orders \\
\hline
MSFT & 220000     & $10^{-16}$           & 20000000 \\
INTC  & 150000     & $10^{-16}$           &  10000000 \\
\hline
\end{tabular}
\label{hypothesis_table}
\end{table}
\par In the following sections we discuss empirical findings for each order type. We find that the well known symmetry of LOB dynamics is fairly well respected. Therefore we will only focus on describing the results from the perspective of the buy side of the LOB.
\subsection{Market orders}
\begin{figure}[h!]
    \begin{subfigure}[t]{0.45\textwidth}
        \centering
        \includegraphics[width=\linewidth]{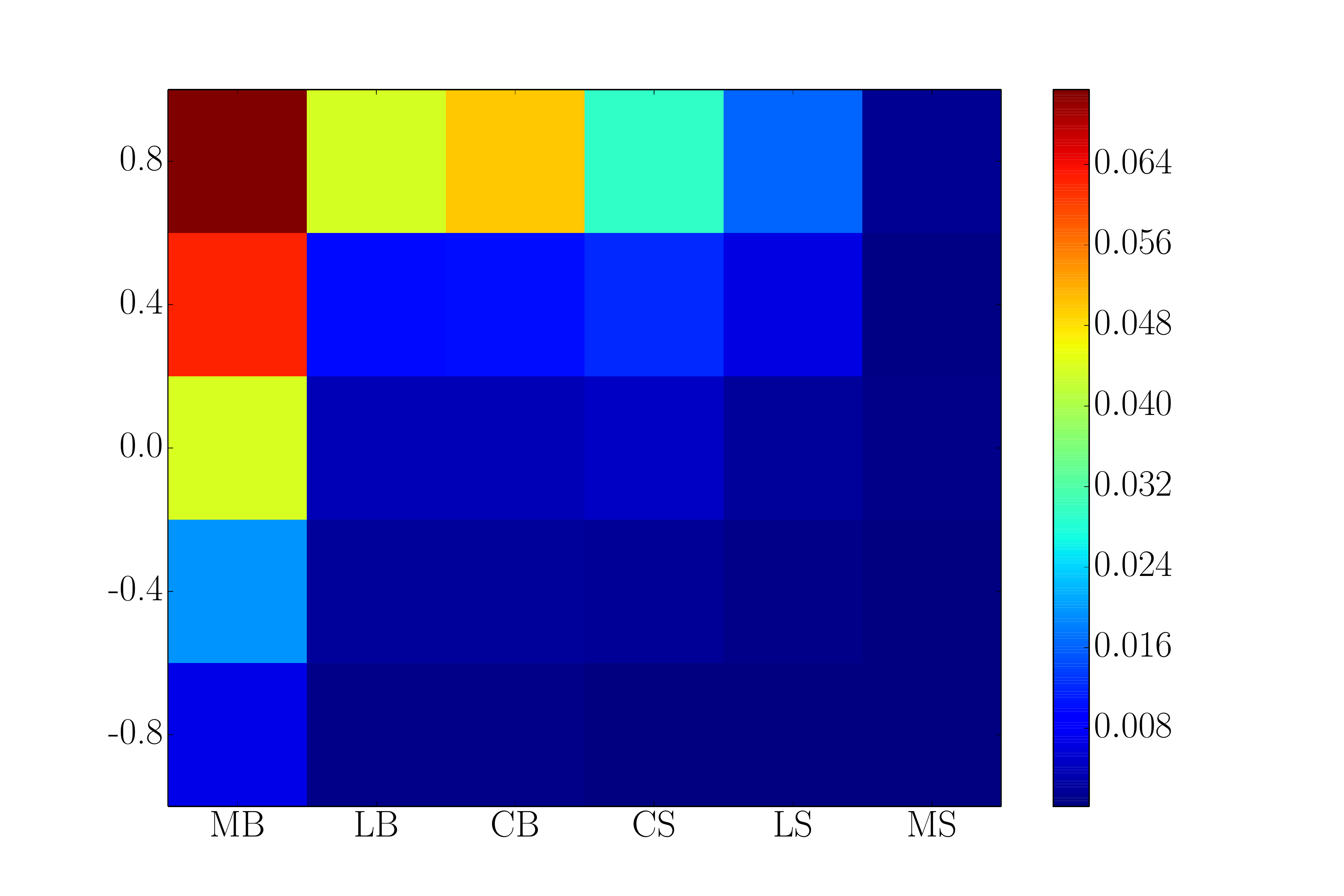} 
        \caption{MSFT} 
        \label{MSFT_heat_MB}
    \end{subfigure}
    	\hfill
    \begin{subfigure}[t]{0.45\textwidth}
        \centering
        \includegraphics[width=\linewidth]{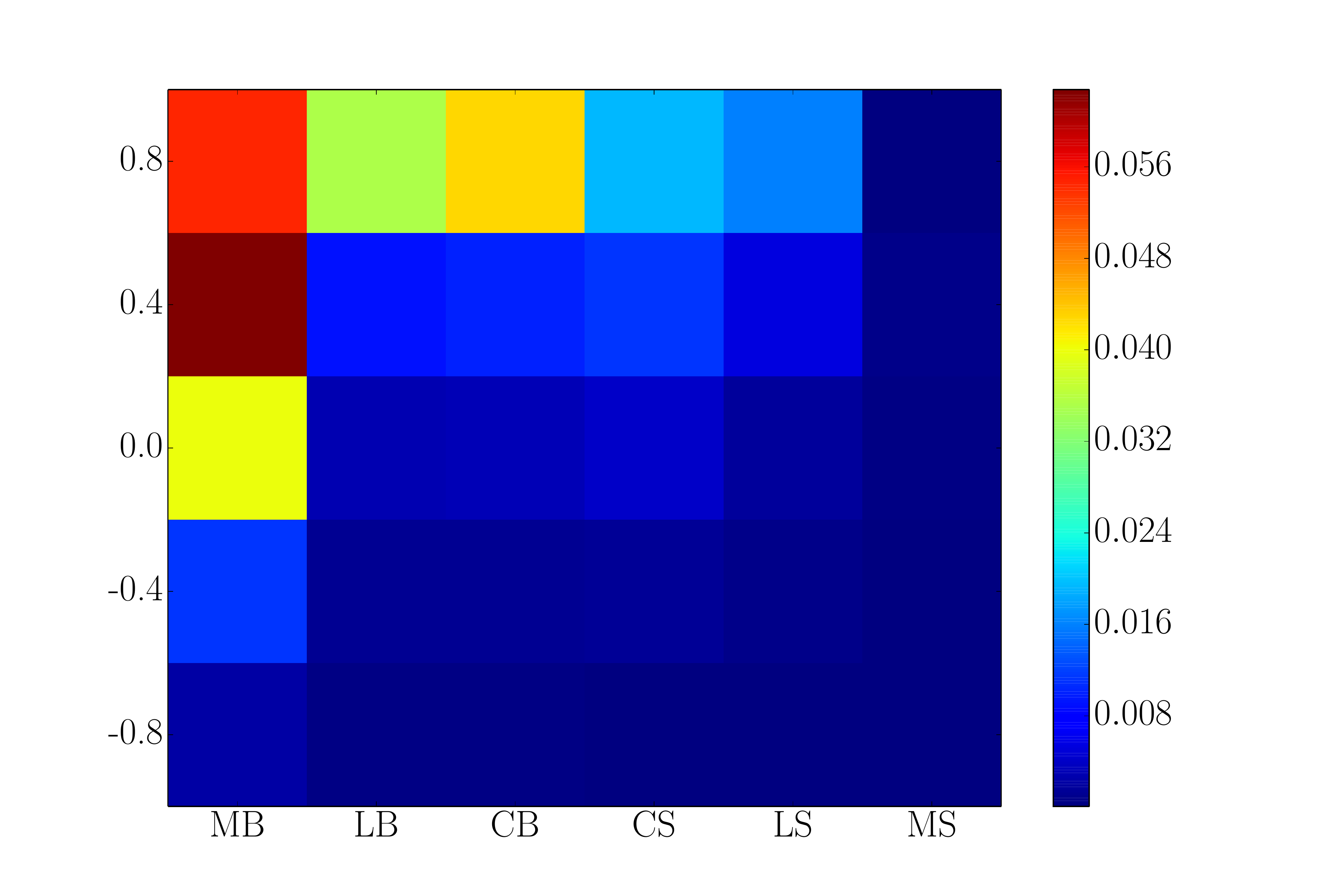} 
        \caption{INTC} 
        \label{INTC_heat_MB}
    \end{subfigure}
    \captionsetup{justification=centering}
    \caption{Heat map of empirical probability of MB orders conditional on the state of the LOB, 
                  x-axis = last order observed, y-axis = imbalance for data on MSFT (left) and INTC (right)}
    \label{heat_MB}
\end{figure}

We start our analysis by considering market buy orders (MB). The heat maps in Fig. \ref{heat_MB} show the probability of observing a MB conditional on the imbalance level and the last order observed. A clear feature is that the probability of observing MBs increases monotonically with imbalance. When the LOB is buy heavy, with a lot more volume at the bid or little volume at the ask, the next mid-price move is more likely to be up. We will discuss this effect in detail in later sections. This leads to orders rushing in to take the remaining liquidity at the ask before the price change. 
\par 
The effect on MBs of the last order observed are significant. Same side market orders have a strong exciting effect at all imbalance levels. This being more pronounced as imbalance increases. Opposite side orders have an equally strong inhibitory effect. Limit and cancellation orders appear to have little influence at all but the highest level of imbalance. These results are to be expected since market orders have been found to be mostly influenced by other market orders Rambaldi, Bacry and Lillo (2016).
\subsection{Limit orders}
\begin{figure}[h!]
    \begin{subfigure}[t]{0.45\textwidth}
        \centering
        \includegraphics[width=\linewidth]{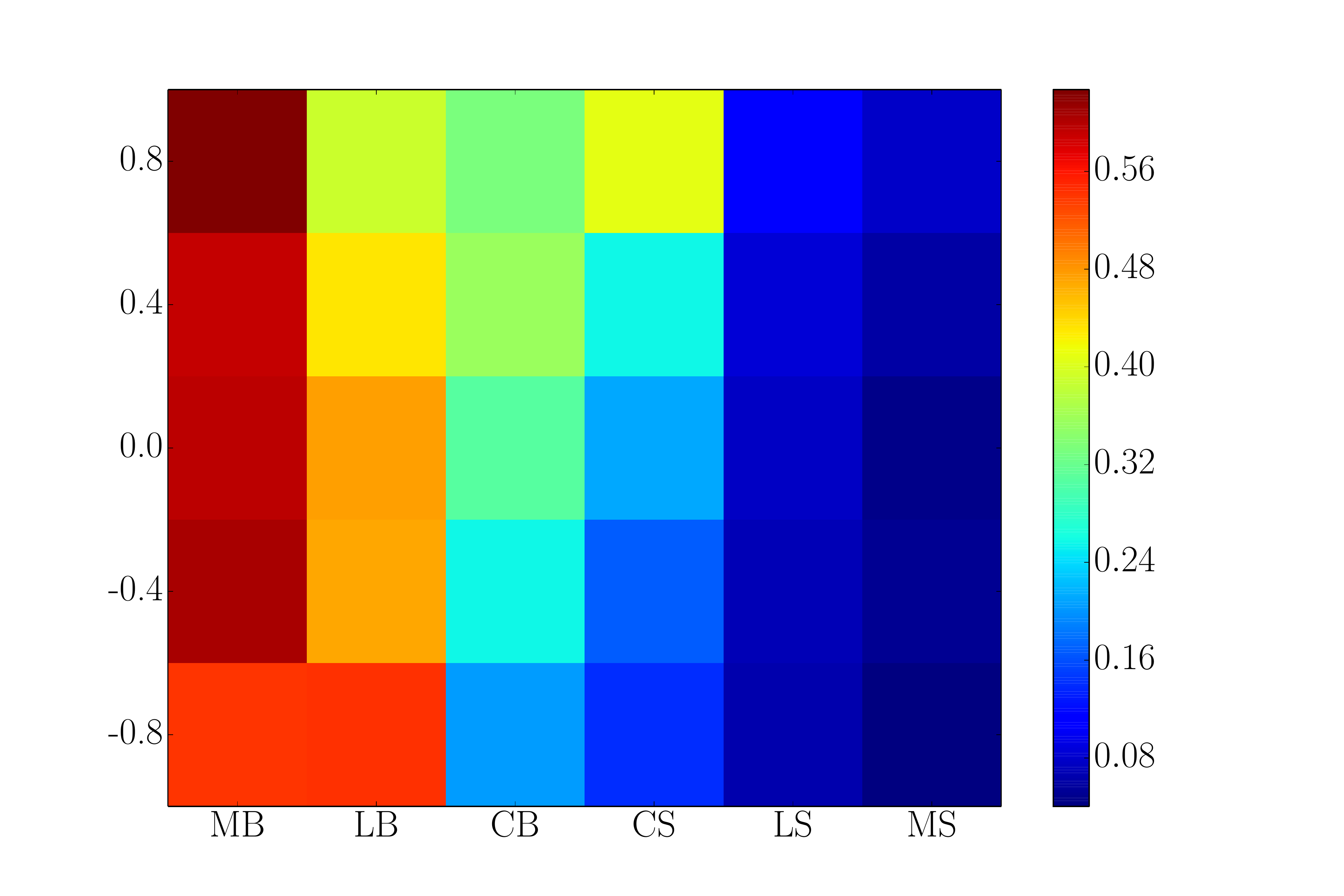} 
        \caption{MSFT} 
        \label{MSFT_heat_LB}
    \end{subfigure}
    	\hfill
    \begin{subfigure}[t]{0.45\textwidth}
        \centering
        \includegraphics[width=\linewidth]{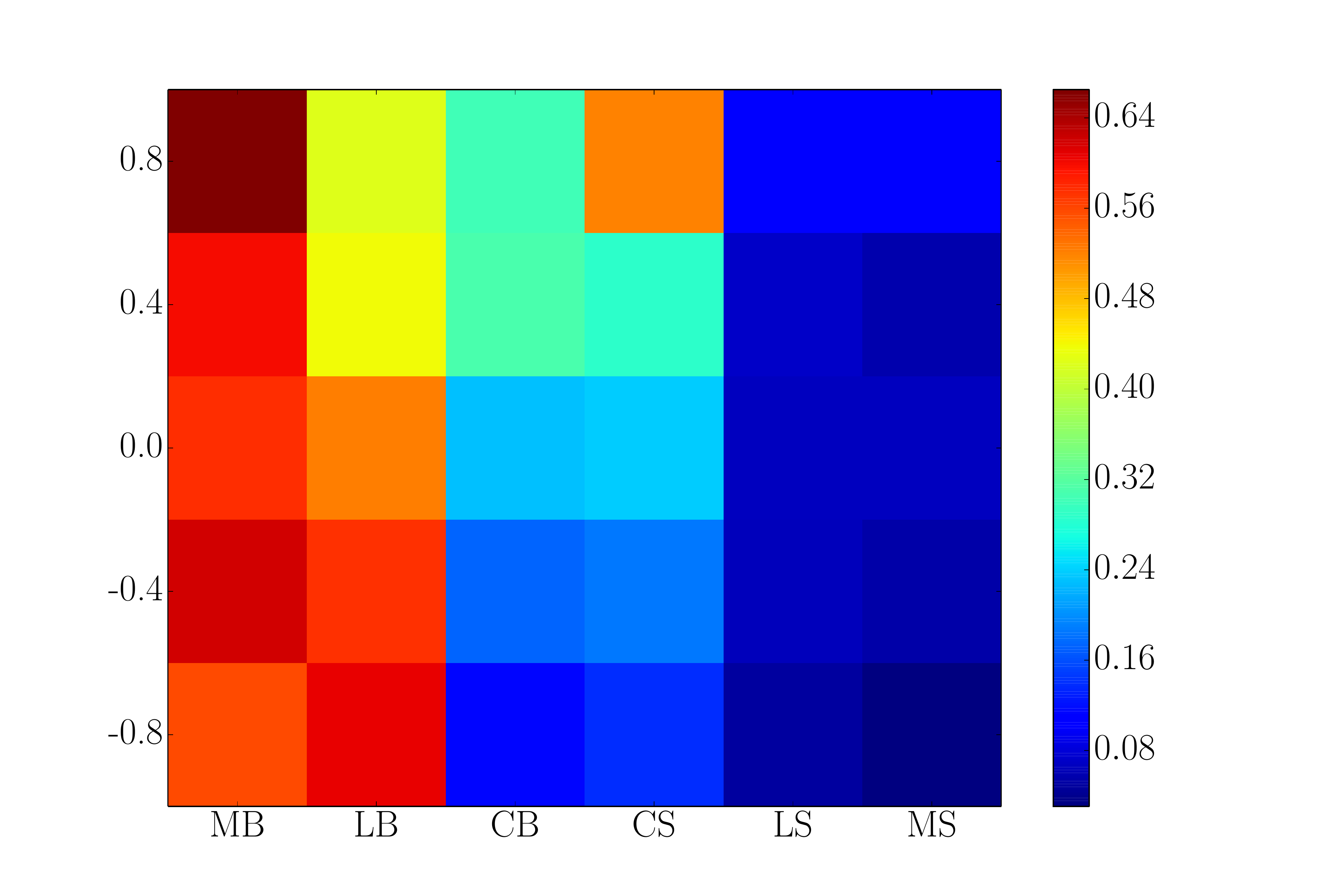} 
        \caption{INTC} 
        \label{INTC_heat_LB}
    \end{subfigure}
    \captionsetup{justification=centering}
    \caption{Heat map of empirical probability of LB orders conditional on the state of the LOB, 
                  x-axis = last order observed, y-axis = imbalance for data on MSFT (left) and INTC (right)}
    \label{heat_LB}
\end{figure}

We now consider limit buy orders (LB). The results are summarized in Fig. \ref{heat_LB}. Similarly as with MBs we see that LB activity increases in general with the imbalance level. However the last order observed appears to be the dominant factor. There is a significant excitation effect from same side LOB activity. Observing a MB or LB significantly increases the probability that a LB will follow, with the influence is of MBs more pronounced than that of LBs. All the effects are amplified by the imbalance level. We also note that there is a symmetric inhibitory effect for opposite side LOB activity. It is remarkable that this inhibitory influence is virtually independent of the imbalance level.  

\subsection{Cancellation orders}
We conclude our analysis by considering cancellation of buy orders (CB). It is worth pointing out that cancellation activity has also been shown to be dependent on the queue position of the order Donelly and Gan (2017). Here we consider the probability of the arrival of a CB at any queue position. As with the previous order types, imbalance is monotonically related to cancellation activity. The effect of orders on the same side is strong, LBs and CBs have an exciting effect on seeing a CB. This is to be expected since a lot of high-frequency trading strategies involve constant queue repositioning. Perhaps surprisingly, other order types seem to have very little effect.  We see that the conditional probability that the next order is CB appears to depend only on imbalance when the last order observed is a MB, CS, LS or MS.
\begin{figure}[h!]
    \begin{subfigure}[t]{0.45\textwidth}
        \centering
        \includegraphics[width=\linewidth]{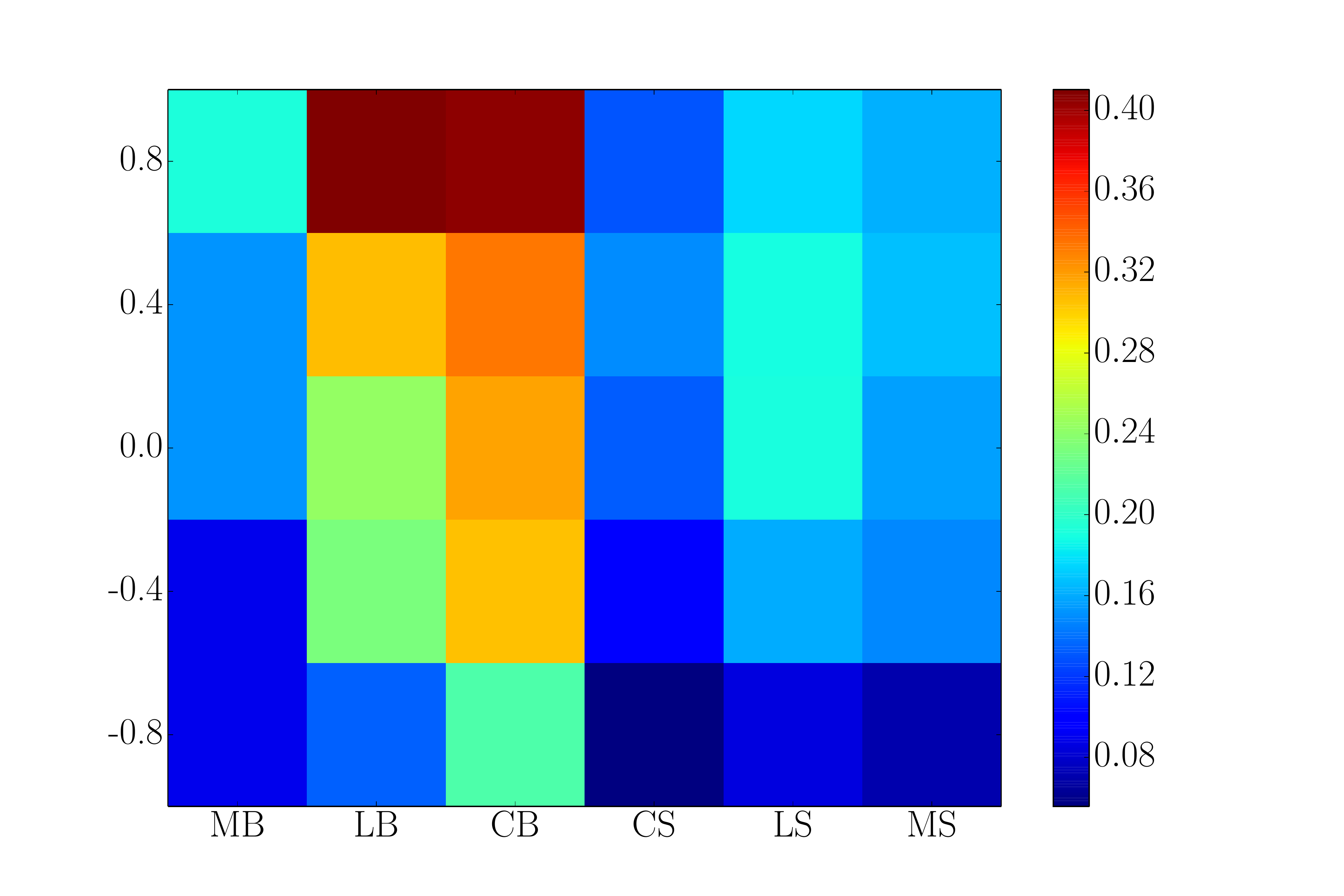} 
        \caption{MSFT} 
        \label{MSFT_heat_CB}
    \end{subfigure}
    	\hfill
    \begin{subfigure}[t]{0.45\textwidth}
        \centering
        \includegraphics[width=\linewidth]{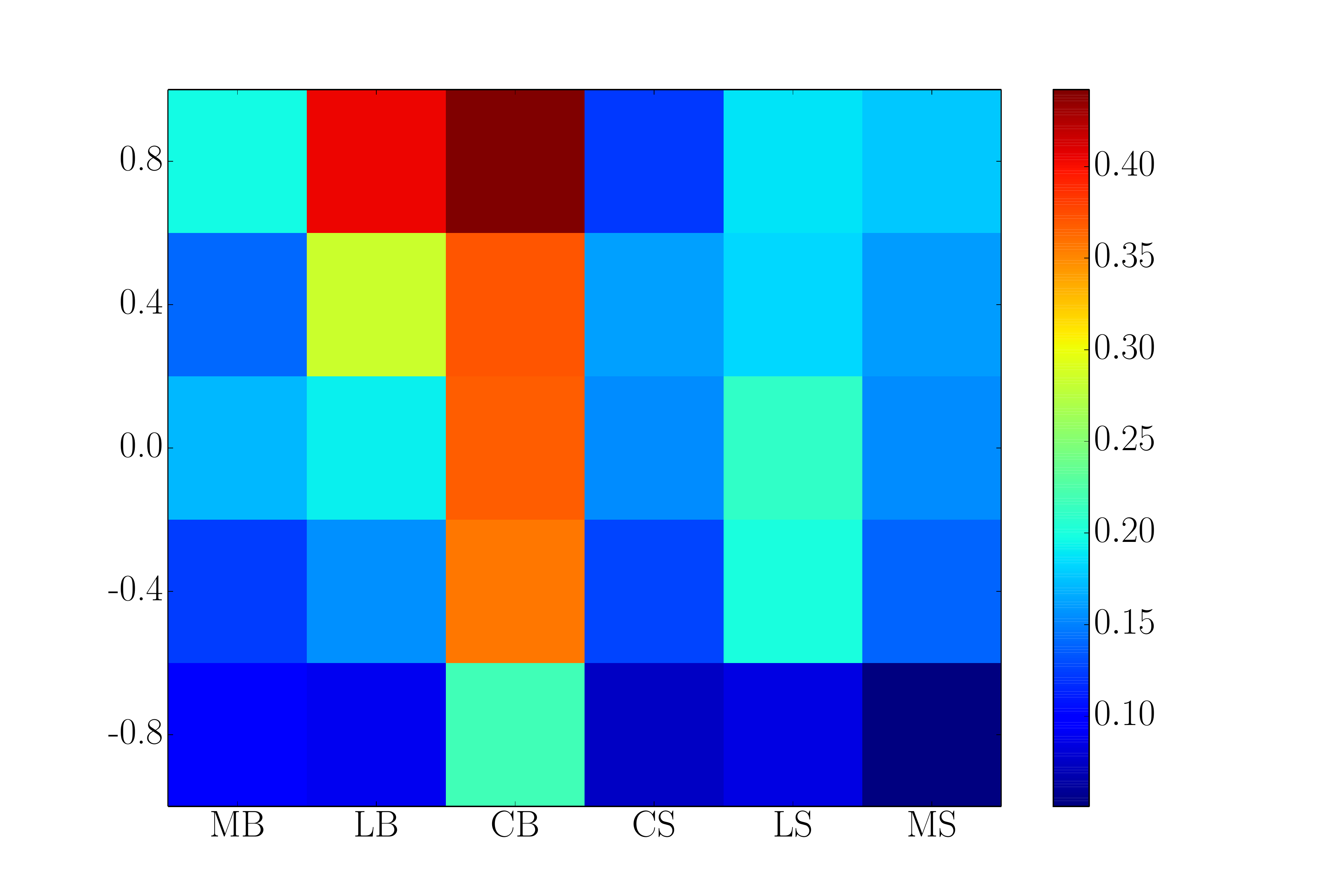} 
        \caption{INTC} 
        \label{INTC_heat_CB}
    \end{subfigure}
    \captionsetup{justification=centering}
    \caption{Heat map of empirical probability of CB orders conditional on the state of the LOB, 
                  x-axis = last order observed, y-axis = imbalance for data on MSFT (left) and INTC (right)}
    \label{heat_CB}
\end{figure}

\subsection{Model}
We now introduce the general framework to describe the LOB dynamics that will be used to optimize high-frequency order submission strategies. As discussed before we will model the dynamics only at the best quotes. In this case the LOB is seen as a 2-dimensional vector which describes the number of limit orders at the bid and ask. To this we will add a categorical variable representing the type of the last order that arrived to the LOB. So the state is summarized by $(V_{b}, V_{a}, e)$. Using the same notation as in the previous section, $V_{b}, V_{a}$ are the numbers of orders at the bid and ask respectively and $e$ denotes the type of the last order observed $e \in E = \{ \text{MB, LB, CB, CS, LS, MS} \}$. 
\par
We model this 3-dimensional process as a discrete time Markov chain with finite states space $\Omega = K^{2}\times E$, where $K$ is the maximal queue size. By our discussion before, the evolution of this Markov chain should depend on the state and history of the LOB. We summarize our assumptions to model the dynamics of this model as follows:
\begin{enumerate}
\item Limit orders and cancellations are of unit size
\item Market orders' sizes depend only on the queue size 
\item The type of the next order arriving depends on the state of the chain
\item When a queue is depleted, the sizes of both the bid and the ask queues are resampled independently according to an empirically estimated queue-size distribution.  
\end{enumerate}
This can be formalized by a 2-step procedure for a given LOB state $(V_{b}, V_{a}, e)$. First the type of the next order arriving has multinomial distribution with parameter $p(V_{b}, V_{a}, e) = p(d, e)$ where $d$ is the discretization of the imbalance given by $V_{b}$ and $V_{a}$. That is, the order arrival type depends on the queue sizes only through their imbalance. Then if the next order is a limit order or cancellation, add or subtract a unit of volume on the corresponding queue and update $e$ accordingly. If the next order is a market order, the size is drawn from a distribution depending on the queue at which it arrives, then update the volume level and last event accordingly.
\par 
If one of the queues is depleted then either the bid goes up or the ask goes down according to a Bernoulli random variable with a stock dependent parameter which measures the probability of consecutive price movements. (A detailed analysis of the mid-price evolution is given in Appendix \ref{imbalance}.) Then after the price move direction has been determined, both queue sizes are resampled independently. The estimation of all the parameters in the model are done via MLE.
\subsection{Relationship with queuing models}
The model we propose is intimately connected to well known continuous time Markov chain LOB models which we will now explain. These models share, as a base assumption, that all orders arrive as Poisson point processes. In the two most well known models, the intensities of these processes are assumed constant Cont et al. (2010), dependent on the LOB shape Huang et al. (2015) and time homogeneous in both. This last assumption allows us to look at these models in terms of the embedded discrete time Markov chain that tracks only the states to which the process jumps.
The transition probabilities of this embedded chain are a simple function of the intensities of the order arrival processes. 
\par For each order type $O\in\{MB,MS,LB,LS,CM,CS\}$, let $\lambda_O(V_{b}, V_{a})$ be the arrival intensity of order type $O$ 
as a function of the LOB state $(V_{b}, V_{a})$ as in Huang et al. (2015). 
These intensity rates are usually estimated via likelihood maximization, e.g. the rate of $LB$ can be estimated by:
\[
\lambda_{LB}(V_{b}, V_{a}) = \frac{1}{\hat{T}(V_{b}, V_{a})}\frac{\hat{N}_{LB}(V_{b}, V_{a})}{\hat{N}(V_{b}, V_{a})}
\]
where $\hat{T}(V_{b}, V_{a})$ is the average time the LOB was in state $(V_{b}, V_{a})$, $\hat{N}_{LB}(V_{b}, V_{a})$ is the total number of LB orders observed in state $(V_{b}, V_{a})$ and $\hat{N}(V_{b}, V_{a})$ is the total number of times state $(V_{b}, V_{a})$ was observed.
\par If all orders are assumed to have constant size 1 we can easily calculate the transition matrix for the embedded chain. As an example:
\[
\mathbb{P}((V_{b}, V_{a}) \to (V_{b}+1, V_{a})) = \frac{\lambda_{LB}(V_{b}, V_{a})}{\lambda(V_{b}, V_{a})}
\]
where $\lambda(V_{b}, V_{a})$ is the sum of the intensities of all six processes. Similarly we can find the transition probabilities to $(V_{b} - 1, V_{a}), (V_{b}, V_{a} + 1) \text{ and } (V_{b}, V_{a} - 1)$. 
If only the embedded chain is of interest, the transition probabilities can be estimated directly from the event counts at each state without first estimating the intensities of the processes. This is the approach we take in our framework. 
\par As a final observation we note that the intensities of the processes can be recovered from the transition probabilities in our model together with the average time spent at each state. 
\section{Optimal order placement}\label{placement}
In this section we consider the problem of how optimally to purchase one share in a LOB. We assume that the trader is able to place and monitor his order reacting to each new order as it arrives to the LOB. The trader's objective is to minimize the cost of purchase given the state of the LOB and its recent history. We allow the trader to submit LOs, cancel them, replace them or send MOs depending on the given market conditions. This discrete optimal control framework has gained popularity, see for example Lehalle and Mounjid (2016). Here we look to build on their framework and extend it. 
\par
In Section \ref{motivation} we motivate our approach for the optimal order placement strategy, in Section \ref{mdp} we introduce the formal framework of Markov decision processes that will be used to derive the optimal order placement strategy and in Section \ref{algo} we present a numerical algorithm that is guaranteed to converge to the optimal solution.
\subsection{Motivation}\label{motivation}
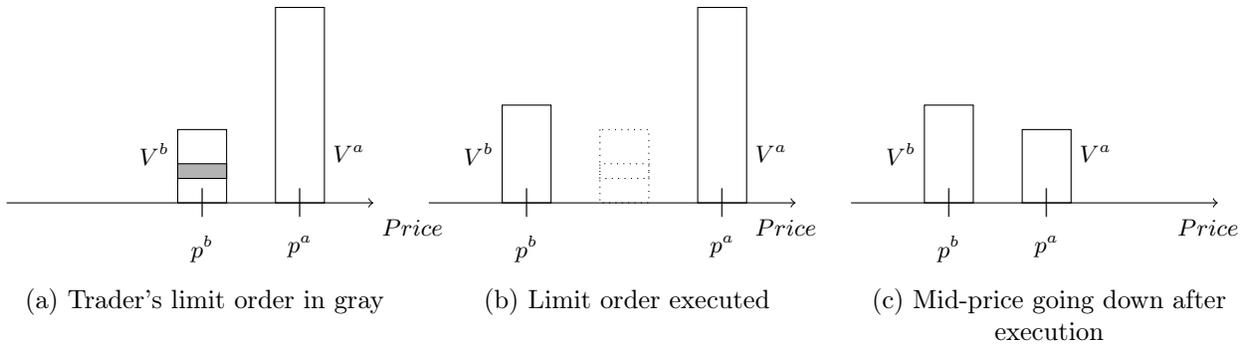
\begin{figure}
\begin{subfigure}[t]{0.32\textwidth}
\begin{tikzpicture}[scale=0.65]
\draw (5.5,-3.5) rectangle (6.5,-5);
\draw [black,fill=gray!60] (5.5,-4.5) rectangle (6.5,-4.2);
\draw (7.5,-1) rectangle (8.5,-5);
\draw [->](2,-5) -- ++(7.5,0);
\draw (6,-5) node[sloped]{$|$};
\draw (8,-5) node[sloped]{$|$};
\footnotesize
\draw (6,-5.5) node[below]{$ p^{b} $} ;
\draw (8,-5.5) node[below]{$ p^{a} $} ;
\draw (5.5,-4) node[left]{$V^{b}$} ;
\draw (8.5,-4) node[right]{$V^{a}$} ;
\draw (9.5,-5.5) node[right]{$Price$} ;
\end{tikzpicture}
\captionsetup{justification=centering}
\caption{Trader's limit order in gray}
\end{subfigure}  
\hfill
\begin{subfigure}[t]{0.32\textwidth}
\begin{tikzpicture}[scale=0.65]
\draw (3.5,-3) rectangle (4.5,-5);
\draw [dotted](5.5,-3.5) rectangle (6.5,-5);
\draw [dotted](5.5,-4.5) rectangle (6.5,-4.2);
\draw (7.5,-1) rectangle (8.5,-5);
\draw [->](2,-5) -- ++(7.5,0);
\draw (4,-5) node[sloped]{$|$};
\draw (8,-5) node[sloped]{$|$};
\footnotesize
\draw (4,-5.5) node[below]{$ p^{b} $} ;
\draw (8,-5.5) node[below]{$ p^{a} $} ;
\draw (3.5,-4) node[left]{$V^{b}$} ;
\draw (8.5,-4) node[right]{$V^{a}$} ;
\draw (8.5,-5.5) node[right]{$Price$} ;
\end{tikzpicture}
\captionsetup{justification=centering}
\caption{Limit order executed}
\end{subfigure}  
\hfill
\begin{subfigure}[t]{0.32\textwidth}
\begin{tikzpicture}[scale=0.65]
\draw (3.5,-3) rectangle (4.5,-5);
\draw (5.5,-3.5) rectangle (6.5,-5);
\draw [->](2,-5) -- ++(7.5,0);
\draw (4,-5) node[sloped]{$|$};
\draw (6,-5) node[sloped]{$|$};
\footnotesize
\draw (4,-5.5) node[below]{$ p^{b} $} ;
\draw (6,-5.5) node[below]{$ p^{a} $} ;
\draw (3.5,-4) node[left]{$V^{b}$} ;
\draw (6.5,-4) node[right]{$V^{a}$} ;
\draw (8.5,-5.5) node[right]{$Price$} ;
\end{tikzpicture}
\captionsetup{justification=centering}
\caption{Mid-price going down after execution}
\end{subfigure}
\caption{Adverse selection risk}
\label{adverse_selection}
\end{figure}

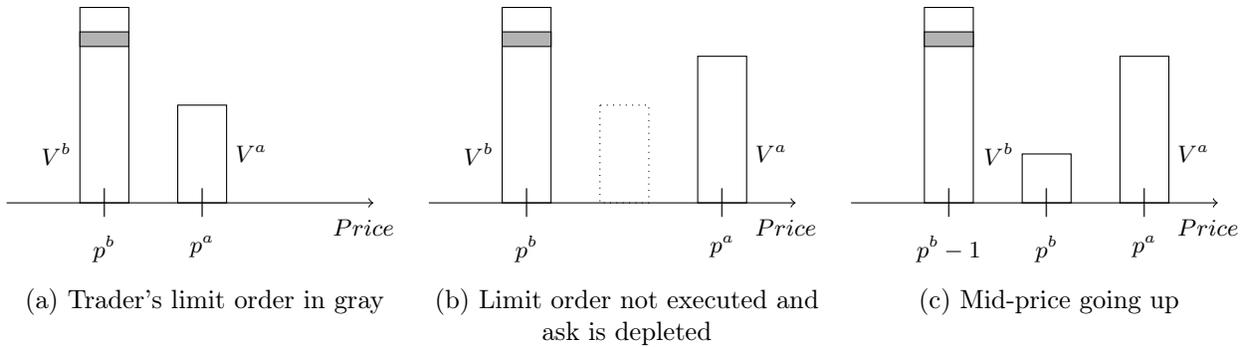
\begin{figure}
\begin{subfigure}[t]{0.32\textwidth}
\begin{tikzpicture}[scale=0.65]
\draw (3.5,-1) rectangle (4.5,-5);
\draw (5.5,-3) rectangle (6.5,-5);
\draw [black,fill=gray!60] (3.5,-1.8) rectangle (4.5,-1.5);
\draw [->](2,-5) -- ++(7.5,0);
\draw (4,-5) node[sloped]{$|$};
\draw (6,-5) node[sloped]{$|$};
\footnotesize
\draw (4,-5.5) node[below]{$ p^{b} $} ;
\draw (6,-5.5) node[below]{$ p^{a} $} ;
\draw (3.5,-4) node[left]{$V^{b}$} ;
\draw (6.5,-4) node[right]{$V^{a}$} ;
\draw (8.5,-5.5) node[right]{$Price$} ;
\end{tikzpicture}
\captionsetup{justification=centering}
\caption{Trader's limit order in gray}
\end{subfigure}  
\hfill
\begin{subfigure}[t]{0.32\textwidth}
\begin{tikzpicture}[scale=0.65]
\draw (3.5,-1) rectangle (4.5,-5);
\draw [black,fill=gray!60] (3.5,-1.8) rectangle (4.5,-1.5);
\draw [dotted](5.5,-3) rectangle (6.5,-5);
\draw (7.5,-2) rectangle (8.5,-5);
\draw [->](2,-5) -- ++(7.5,0);
\draw (4,-5) node[sloped]{$|$};
\draw (8,-5) node[sloped]{$|$};
\footnotesize
\draw (4,-5.5) node[below]{$ p^{b} $} ;
\draw (8,-5.5) node[below]{$ p^{a} $} ;
\draw (3.5,-4) node[left]{$V^{b}$} ;
\draw (8.5,-4) node[right]{$V^{a}$} ;
\draw (8.5,-5.5) node[right]{$Price$} ;
\end{tikzpicture}
\captionsetup{justification=centering}
\caption{Limit order not executed and ask is depleted}
\end{subfigure}  
\hfill
\begin{subfigure}[t]{0.32\textwidth}
\begin{tikzpicture}[scale=0.65]
\draw (3.5,-1) rectangle (4.5,-5);
\draw [black,fill=gray!60] (3.5,-1.8) rectangle (4.5,-1.5);
\draw (5.5,-4) rectangle (6.5,-5);
\draw (7.5,-2) rectangle (8.5,-5);
\draw [->](2,-5) -- ++(7.5,0);
\draw (4,-5) node[sloped]{$|$};
\draw (6,-5) node[sloped]{$|$};
\draw (8,-5) node[sloped]{$|$};
\footnotesize
\draw (4,-5.5) node[below]{$ p^{b}-1$};
\draw (6,-5.5) node[below]{$ p^{b} $} ;
\draw (8,-5.5) node[below]{$ p^{a} $} ;
\draw (5.5,-4) node[left]{$V^{b}$} ;
\draw (8.5,-4) node[right]{$V^{a}$} ;
\draw (8.5,-5.5) node[right]{$Price$} ;
\end{tikzpicture}
\captionsetup{justification=centering}
\caption{Mid-price going up}
\end{subfigure}
\caption{Non-execution risk}
\label{non_execution}
\end{figure}

When interacting directly with the LOB via LOs and MOs market participants face two main short term risks.
\begin{itemize}
\item Adverse selection risk shown in Fig. \ref{adverse_selection}. This occurs when the trader has an active LO in the book, the order is executed and then the price moves against him, i.e. a LB (LS) is executed and then the price moves down (up). Ideally the trader would have cancelled his limit order and waited after the price move. This would allow the trader to benefit from the decrease (increase) by buying (selling) at a lower (higher) price. 
\item
Second is non-execution risk shown in Fig. \ref{non_execution}. This occurs when the trader has an active LO in the book, the order isn't executed and then the price moves against him, i.e. a LB (LS) is not executed and then the price moves up (down). In this situation the trader should have cancelled his limit order and submitted a MO before the price move. 
\end{itemize}

\par

\par There has been considerable work done to derive optimal trading strategies using LOB information.  We offer two particular extensions which we will show help to reduce the risks of both adverse selection and non-execution. First, we explicitly model how order flow history impacts LOB dynamics and measure its affect on optimal trading.  Second, we make use of all of the order types available to traders in real markets and show how LOs, COs and MOs are all essential to optimal trading.

\par Our contributions in this problem can be summarized as follows.  First the dynamics that we assume for the evolution of the LOB incorporate the influence of both the state of the LOB and recent historical order flow, which is more realistic than existing models. Additionally, our model accounts for the instantaneous impact of the trader's own orders in the subsequent evolution of the LOB. Second, our study of the LOB imbalance signal allows us to define the objective function in the trading problem more precisely by understanding the time horizon in which the imbalance signal contains useful information. Finally and most importantly, we allow our trading strategy to use all order types depending on the market conditions. This allows the trader to avoid the two risks described above and obtain the optimal execution price.

\subsection{Markov decision process formulation}\label{mdp}

In this section we briefly introduce the general framework of Markov decision processes and then show how we can formalize and solve the optimal order placement problem. For a more general treatment on the subject see for example Kochenderfer (2015). 
\par
Let $M = (S,T)$ be a Markov chain, where $S$ is the countable state space and $T$ its transition matrix. A discrete Markov decision process (MDP) is a Markov chain where an agent is allowed to take actions that affect the dynamic evolution of the system. Formally, it is a Markov chain with three additional components, a finite set $A$ of all possible actions, a function $a\colon S \to 2^{A}$ that determines the actions admissible at each state and a reward function $R\colon S\times A \to \mathbb{R}$. The transition matrix $T$ is allowed to depend on the actions taken by the agent. A policy is a function $\pi \colon S \to A$ that maps each state $s$ into an admissible action $a(s)$. The expected reward  $U \colon S \to \mathbb{R}$ for the agent with a given policy $\pi$ is given by:
\begin{equation}\label{value_def}
U(s, \pi) = \mathbb{E}\left[ \left.\sum_{i=1}^{\infty} R(s_{i+1} | s_{i}, \pi(s_{i}))\right| s_{0} = s \right]
\end{equation}
\par The problem facing the agent is finding the policy that maximizes the value function. That is, for a state $s$ the agent wants to know which admissible action $a(s)$ maximizes his expected reward given that he is in state $s$. Under mild regularity conditions for the MDP, the optimal policy $\pi^{*}$ exists and its given by:
\begin{equation}\label{optimal_policy_def}
\pi^{*}(s) = \max_{a \in a(s)} \left[ R(s,a) + \sum\limits_{s' \in N(s)}T(s'|s,a) U_{k}(s', \pi^{*}) \right], \text{ for every } s \in S
\end{equation}
where $N(s)$ is the set of states reachable from state $s$ after taking action $a$. The expected reward under the optimal policy is called the value function and is given by $U(s) = U(s,\pi^{*})$. From the value function it is trivial to recover the optimal policy. 
\subsubsection{Single period optimal order placement}
We want to show how the problem of optimal purchase of 1 share can be represented as finding the optimal policy in a MDP. We start with a simplified version of the problem where the trader only has until the mid-price movement to buy the share. Our assumptions can be summarized as follows:
\begin{itemize}
\item The spread is equal to 1 and doesn't change throughout the trading period.
\item The trader can have at most 1 LO active in the book.
\item If the share hasn't been bought and the mid-price moves the trader must immediately submit a MB. 
\item Volume at the bid and ask is bounded by some constant $K$.
\end{itemize}
Next we specify each of the components of the MDP:
\begin{itemize}
\item The state space $S$ is given by the orders in front and including the traders' order, the number of orders after, the orders on the opposite side, the type of the last order observed $E = \{MB,MS,LB,LS,CB,CS\}$ and a categorical variable $I$ describing whether the trader has no active LO (a), active LO (b), bought the share with a LO (c) or bought the share with a MO (d). We will denote the state of the process at time $t$ by $s_{t} =(V^{b,1}_{t}, V^{b,2}_{t}, V^{a}_{t}, e_{t}, i_{t})$.
\item For a given state $s_{t} =(V^{b,1}_{t}, V^{b,2}_{t}, V^{a}_{t}, e_{t}, i_{t})$, following the notation used before we reduce it to $(D_{t},e_{t})$, where $D_{t}$ is the discretized imbalance of the LOB. The transition probabilities at state $s_{t}$ depend on the probability of arrival of each order type given the reduced state $(D_{t}, e_{t})$ as described in Section \ref{analysis1}. Again we assume that all LOs and COs are of size 1 and the sizes of MOs depend only on the size of the queue at which they arrive. We also define absorbing states to be when either the bid or the ask has 0 volume. This determines the transition matrix $T$ completely. 
\item The set of actions $A$ available to the trader are either to wait or to submit one of the three order types MO, LO and CO. Which order types are admissible clearly depends on the state $s_{t}$ and our assumptions. If the trader has no order active in the LOB $(i_{t} = a)$, then he can submit a LO, a MO or wait. If he has an active LO $(i_{t} = b)$ he can submit a CO or wait. If he has already purchased the share $(i_{t} = c \text{ or } d)$ then his only available action is waiting. Every time the trader submits an order, the queue sizes and $e_{t}$ change, so the trader's own actions are reflected in the subsequent evolution of the LOB. 
\item The reward function $R$ is taken to be the the mid-price after the bid or the ask is depleted and refilled minus the acquisition price. The only states where the reward is non-zero are the absorbing states of the chain. We can easily add a cost/rebate into the reward function, for simplicity we omit these adjustments. This choice of reward function is motivated by our analysis in Appendix \ref{imbalance} which shows that LOB imbalance has short term predictive power, i.e., only until the next two mid-price change. The reward function can be summarized by:

\begin{table}[h!]
\caption{Reward function in ticks}
\centering
\begin{tabular}{c | c | c }
Bought with & Mid-price up & Mid-price down  \\
\hline
MO & $0.5$ & -$1.5$ \\
\hline
LO & $1.5$ & -0.5 \\
\hline
MO after mid-price move & -$0.5$ & -0.5
\end{tabular}
\end{table}

\end{itemize}

\subsubsection{Multi period optimal order placement}
With some simple modifications the framework introduced before can be extended to allow the share to be purchased over $M$ mid-price changes to which we'll refer as periods. The assumptions are similar
\begin{itemize}
\item The trader can only have one active LO.
\item If the share hasn't been bought at the end of the $M$ periods the trader must submit a MB.
\item Volume at the bid and the ask are bounded by some constant $K$.
\end{itemize}
The components of the MDP in this case are:
\begin{itemize}
\item The state space $S$ is extended to include the number $m_{t}$ of periods remaining. We also add an additional value (e) to the categorical variable $I$ to indicate that the share was purchased on a previous time period.
\item The transition matrix $T$ remains mostly unchanged. In previous absorbing states, i.e. once a queue was depleted, the system would transition as follows: time periods remaining $m_{t}$ would decrease by one unless $m_{t} = 1$ in which case we would reach an absorbing state, an active LO that wasn't executed is cancelled and the volume at both the bid and the ask are resampled from the empirical distribution of the queue sizes.
\item The set of actions $A$ remains unchanged.
\item For the reasons outlined in Appendix \ref{imbalance}, the reward function $R$ is again taken to be the difference between (i) the mid-price after the immediate next bid or ask queue depletion and refill  and (ii) the acquisition price. 
\end{itemize}

\subsection{Algorithms and convergence}\label{algo}
In this section we present algorithms to solve the optimal order placement problem. There are many algorithms to derive the optimal policy on an MDP. Due to its simplicity and speed of convergence we propose to use a value iteration algorithm. Once the value function is calculated, the optimal policy can be easily derived by taking the action at each state that matches the expected reward given by the value function. The value function for the single period problem can be estimated by Algorithm 1 detailed below: 
\begin{algorithm}[h!]
\label{algo1}
\begin{algorithmic}[1]
\caption{Value iteration algorithm}
\Function{ValueIteration}{}
\State $k \leftarrow 0$
\State $U_{0}(s) \leftarrow -\infty$ for all states $s$
\Repeat
\State $U_{k}(s) \leftarrow \max\limits_{a \in a(s)} \left[R(s,a) + \sum\limits_{s' \in N(s)}T(s'|s,a) U_{k}(s')\right]$ for all states $s$
\State $k \leftarrow k+1$
\Until{convergence}
\State \Return{$U_{k}$}
\EndFunction
\end{algorithmic}
\end{algorithm}
\par Once the value function $U$ has been computed, we can extract the optimal action for each state $s$ by choosing $a \in a(s)$ that maximizes:
\begin{equation}
R(s,a) + \sum\limits_{s' \in N(s)}T(s'|s,a) U_{k}(s')
\end{equation}
\par For the multiple period problem the same algorithm above would converge to the value function. However given the structure of the state space a simple modification increases the convergence rate significantly. The value function for states with $m$ periods remaining only depends on the value function at states with $m$ and $m-1$ periods remaining. This structure lends itself to a dynamic programming type reorganization of the iterations in Algorithm 1. We can compute the value function in waves of states increasing in the number of periods remaining. The detailed Algorithm 2 is shown below.
\begin{algorithm}[h!]
\begin{algorithmic}[1]
\caption{Dynamic value iteration algorithm}
\Function{DynamicValueIteration}{}
\State $j \leftarrow 0$
\State $U_{0}(s) \leftarrow -\infty$ for all states $s$
\Repeat
\State $k \leftarrow 0$
\Repeat
\State $U_{k}(s) \leftarrow \max\limits_{a \in a(s)} \left[R(s,a) + \sum\limits_{s' \in N(s)}T(s'|s,a) U_{k}(s')\right]$ for all states $s$ with $j$ periods left
\State $k \leftarrow k+1$
\Until{convergence}
\State $j \leftarrow j+1$
\Until{$j = M$}
\State \Return{$U_{k}$}
\EndFunction
\end{algorithmic}
\end{algorithm}
\par Convergence of both algorithms is guaranteed by the results proved in Hult and Kiessling (2010). The only thing we need to verify is that under the universe of policies we consider, all of them reach an absorbing state in a finite number of transitions with probability 1. This follows immediately from the fact that every state in our Markov chains is connected to an absorbing state, i.e. from every state there is positive probability to reach an absorbing state.  
\section{Description and performance of the optimal strategy}\label{strategy}
In this section we describe the characteristics of the optimal strategy for the order placement problem with LOB parameters estimated from MSFT sample data. We point out in which states of the LOB MOs, LOs and COs are optimal. Then we also present a simulation experiment, where we compare our strategy with other common ones in the literature. With this experiment we show the value of incorporating all order types into trading strategies.
\subsection{Market orders}
\begin{figure}[h!]
    \begin{subfigure}[t]{0.45\textwidth}
        \centering
        \includegraphics[scale = 0.6]{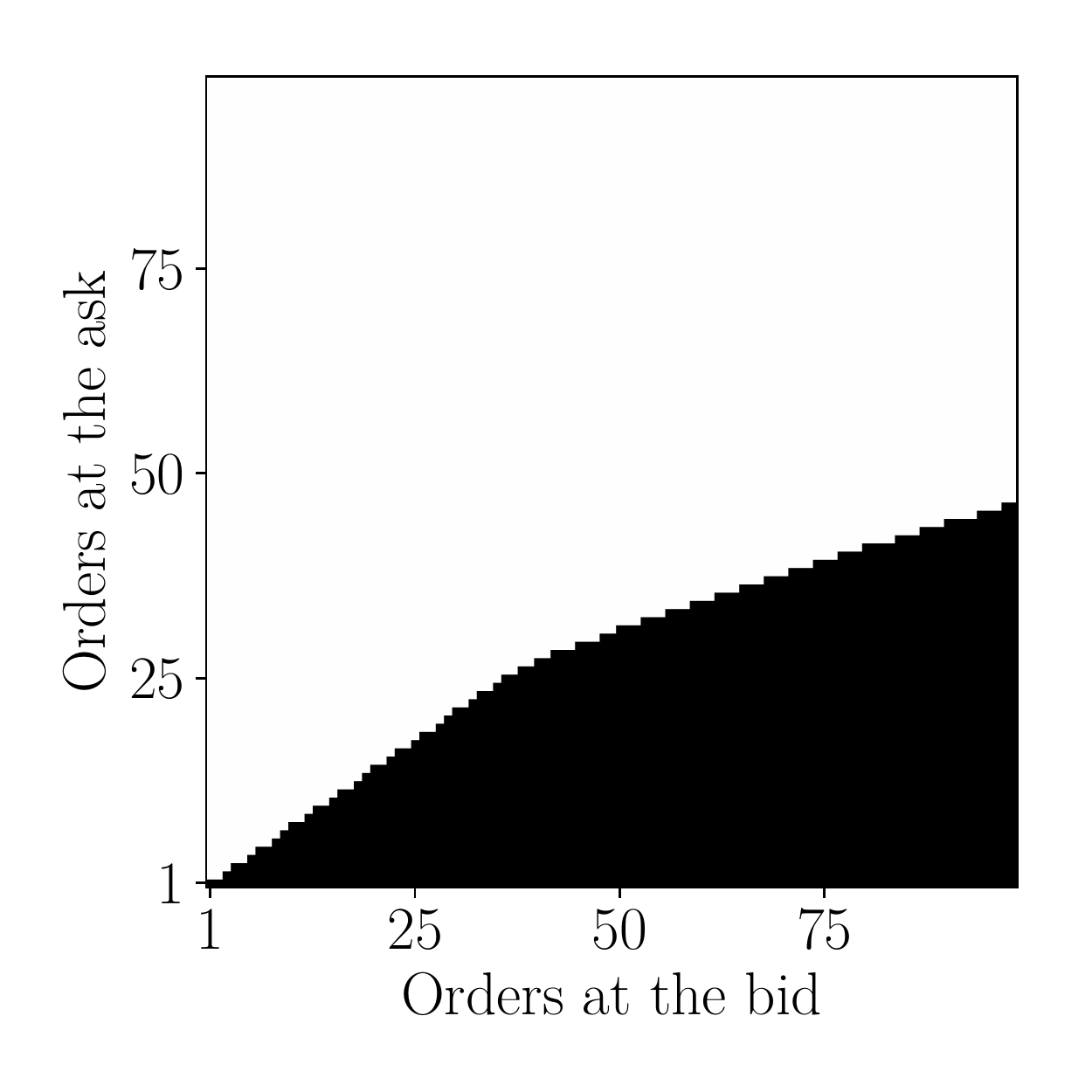} 
        \caption{1 Period remaining} 
        \label{MO_1_period}
    \end{subfigure}
    	\hfill
    \begin{subfigure}[t]{0.45\textwidth}
        \centering
        \includegraphics[scale = 0.6]{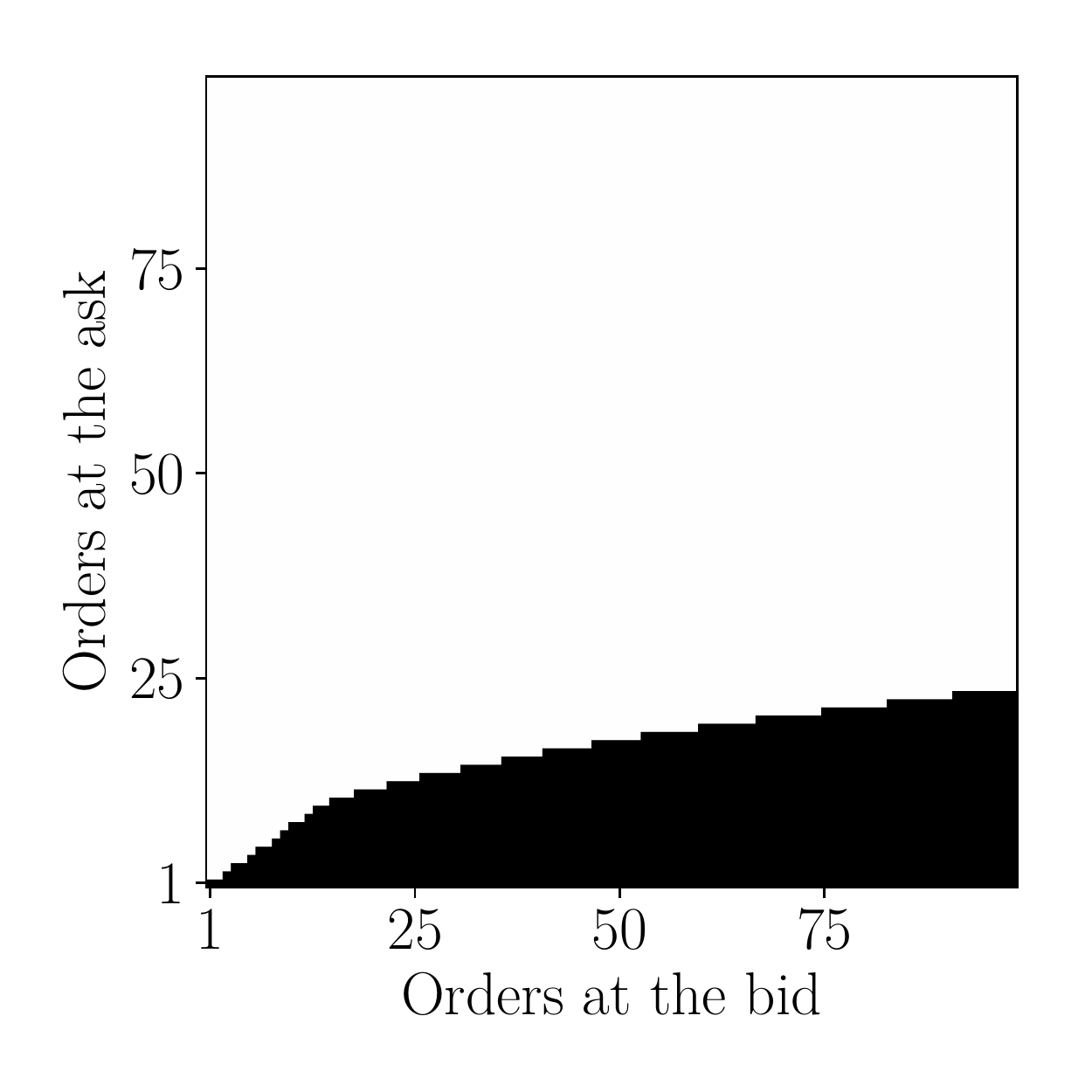} 
        \caption{10 Periods remaining} 
        \label{MO_10_period}
    \end{subfigure}
    \captionsetup{justification=centering}
    \caption{Regions of the LOB where MO submission is optimal (black) for 2 different time horizons for a trader looking to buy}
    \label{mo_region}
\end{figure}

First we will discuss the LOB regions where MOs are optimal. In Fig. \ref{mo_region} the states of the LOB in black indicate where the strategy submits MOs. We assume that the last order was a LB. For other order types, the region changes slightly so we don't include them here. The left panel shows this region when there is 1 period remaining to purchase the share and the right panel shows the region when there are 10 periods remaining to purchase the share.
\par 
There are several characteristics of these plots worthy of discussion. First it is clear that the strategy is more aggressive when there is less time to purchase the share. When comparing the regions in the two panels it can be seen that the MO submission region is much larger on the left plot. This is a desirable and intuitive quality of the solution. The strategy is more patient when there is more time remaining. When there is only 1 period remaining and little volume on both sides, the strategy submits MOs as soon as there is even a slight LOB imbalance. On the other hand, when there are multiple periods remaining the strategy only submits MOs under extreme imbalance. 
\par 
Second, that the optimal strategy submits MOs only when the LOB is imbalanced matches the observed order flow, i.e. MO submission is much higher when the LOB is imbalanced. We can explain this intuitively by noting that if imbalance is an indicator of the future mid-price, then it is natural to aggressively take the remaining volume before the likely price move. Alternatively a LO placed under this scenario of imbalance would face significant non-execution risk, i.e. it is unlikely to be executed before a mid-price change.
\subsection{Limit and cancellation orders}
\begin{figure}[h!]
    \begin{subfigure}[t]{0.45\textwidth}
        \centering
        \includegraphics[scale = 0.6]{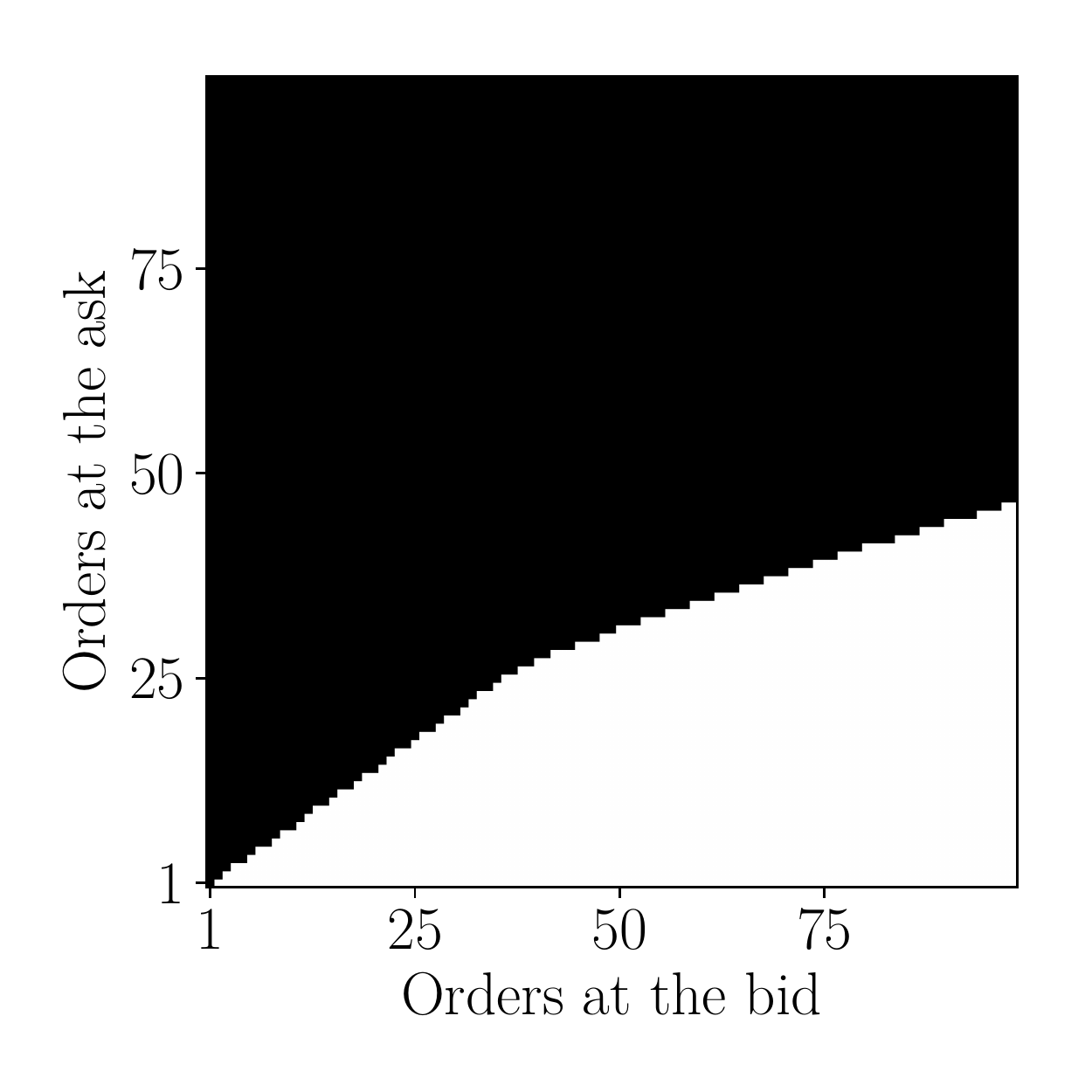} 
        \caption{1 Period remaining} 
        \label{LO_1_period}
    \end{subfigure}
    	\hfill
    \begin{subfigure}[t]{0.45\textwidth}
        \centering
        \includegraphics[scale = 0.6]{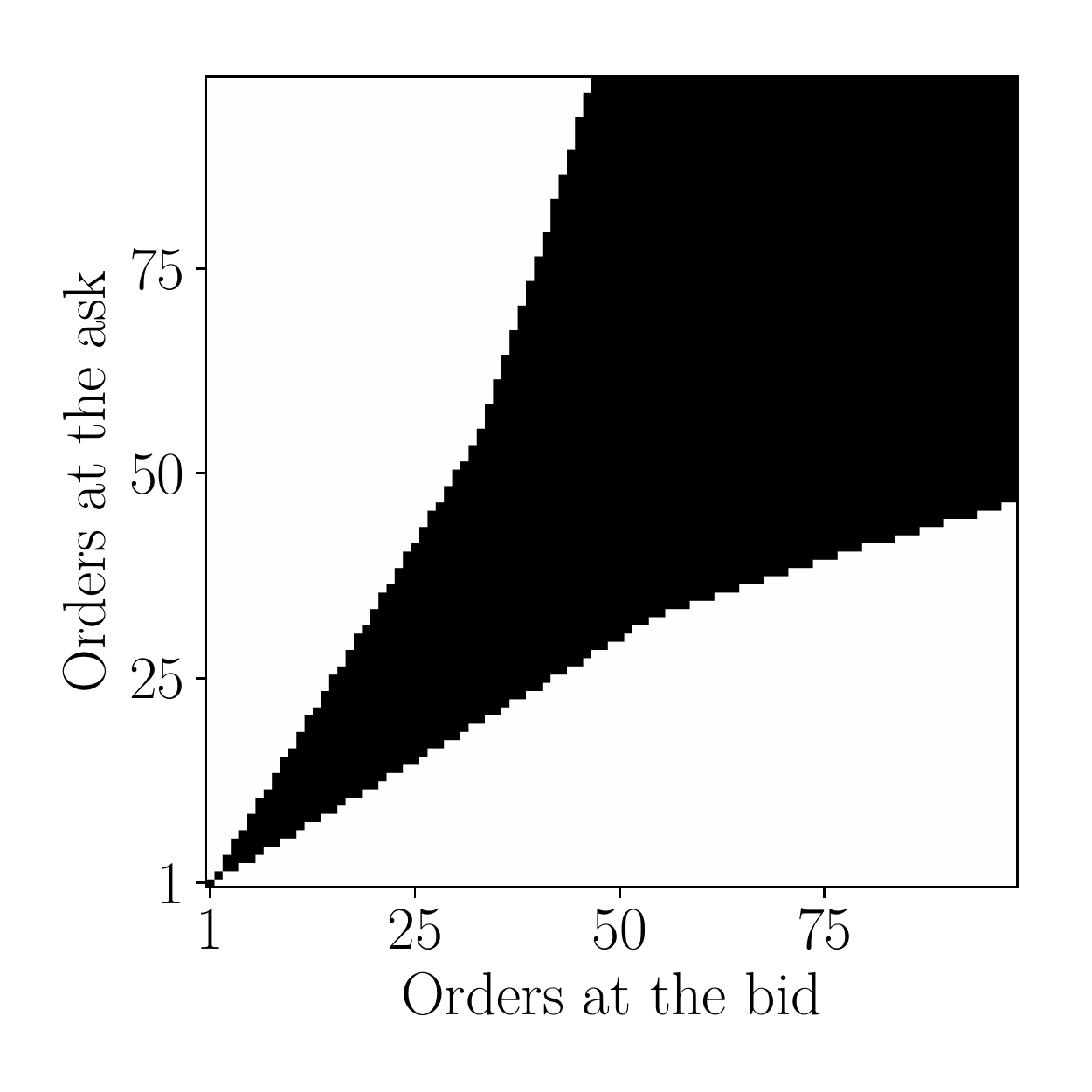} 
        \caption{10 Periods remaining} 
        \label{LO_10_period}
    \end{subfigure}
    \captionsetup{justification=centering}
    \caption{Regions of the LOB where LO submission is optimal (black) for 2 different time horizons for a trader looking to buy}
    \label{lo_region}
\end{figure}

To complete the description of the strategy we now describe the regions where placing a LO is optimal. We show the regions in black for 2 different time horizons in Fig. \ref{lo_region}, the left panel shows the region for 1 period remaining and the right panel for 10 periods remaining. Similar to the MO case the results are very intuitive. As more time remains the strategy is more selective for when to place a LO. When there is more time available the strategy only places LO when the LOB is balanced. The explanation for this last observation is simple. When the imbalance is positive there is high non-execution risk for the LO and when imbalance is negative adverse selection risk is high. 
\par
\begin{figure}[h!]
    \begin{subfigure}[t]{0.45\textwidth}
        \centering
        \includegraphics[scale = 0.6]{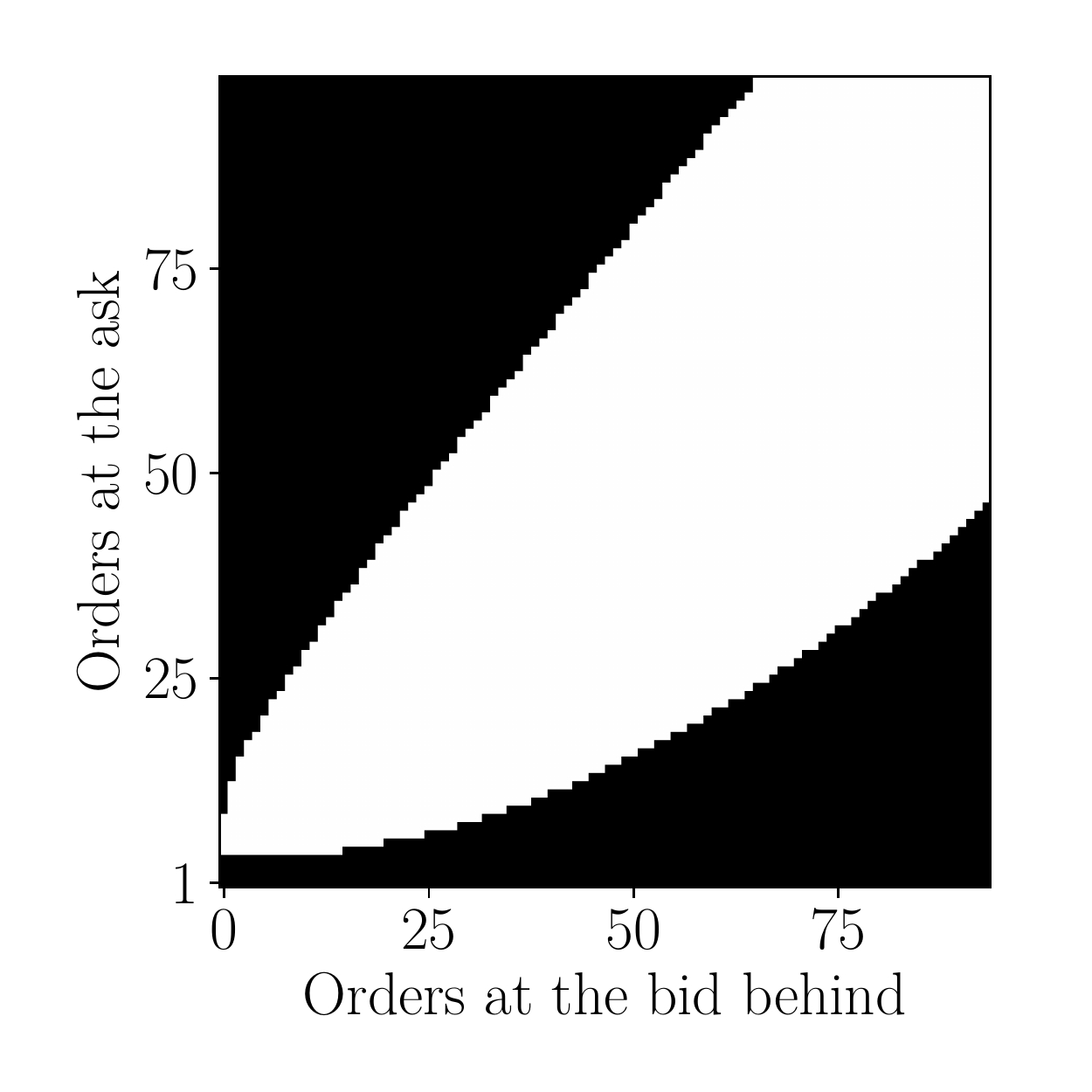} 
        \caption{Order in position 5 at the bid} 
        \label{CO_10_period_5_pos}
    \end{subfigure}
    	\hfill
    \begin{subfigure}[t]{0.45\textwidth}
        \centering
        \includegraphics[scale = 0.6]{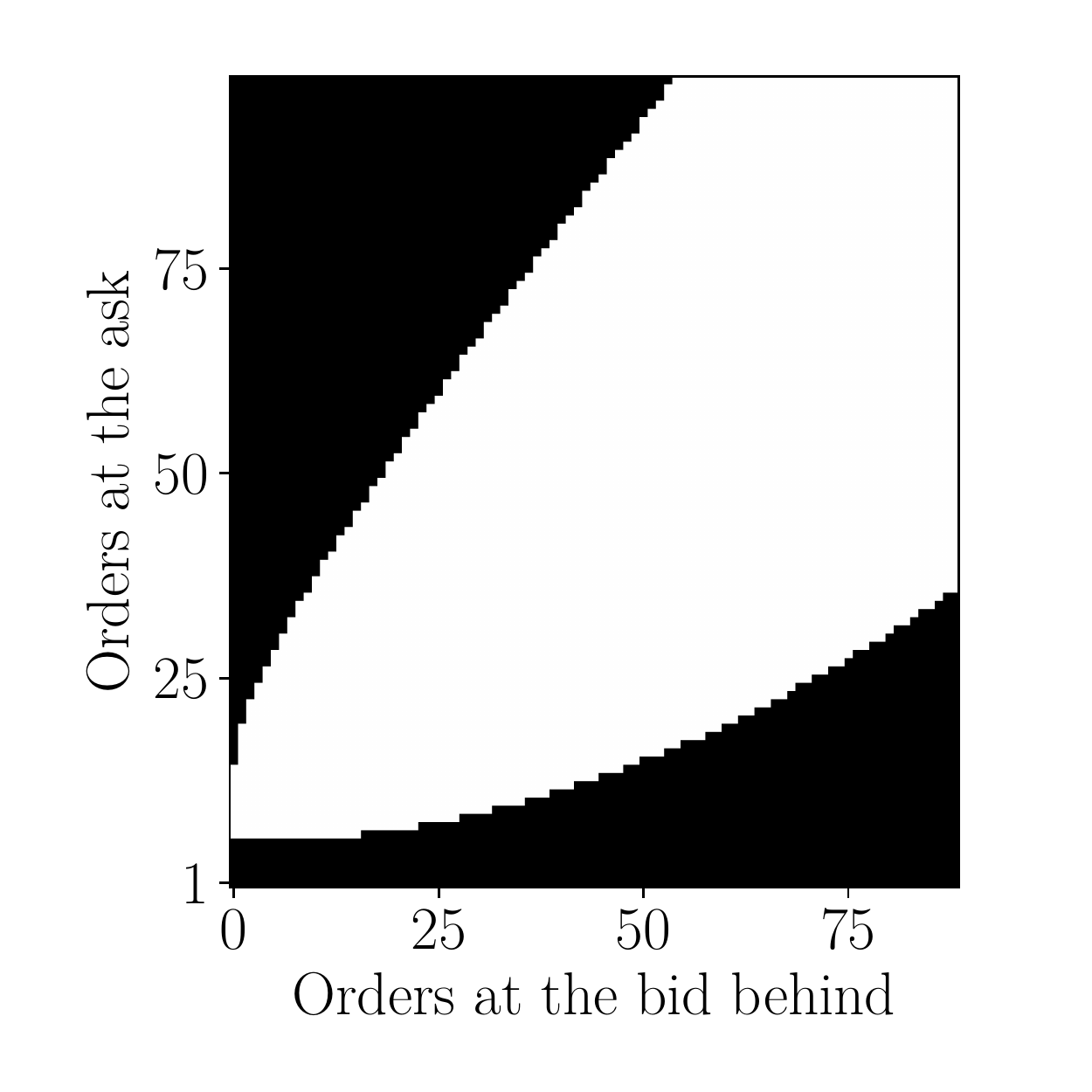} 
        \caption{Order in position 10 at the bid} 
        \label{CO_10_period_10_pos}
    \end{subfigure}
        \vspace{1cm}
        \hfill
     \begin{subfigure}[t]{0.45\textwidth}
        \centering
        \includegraphics[scale = 0.6]{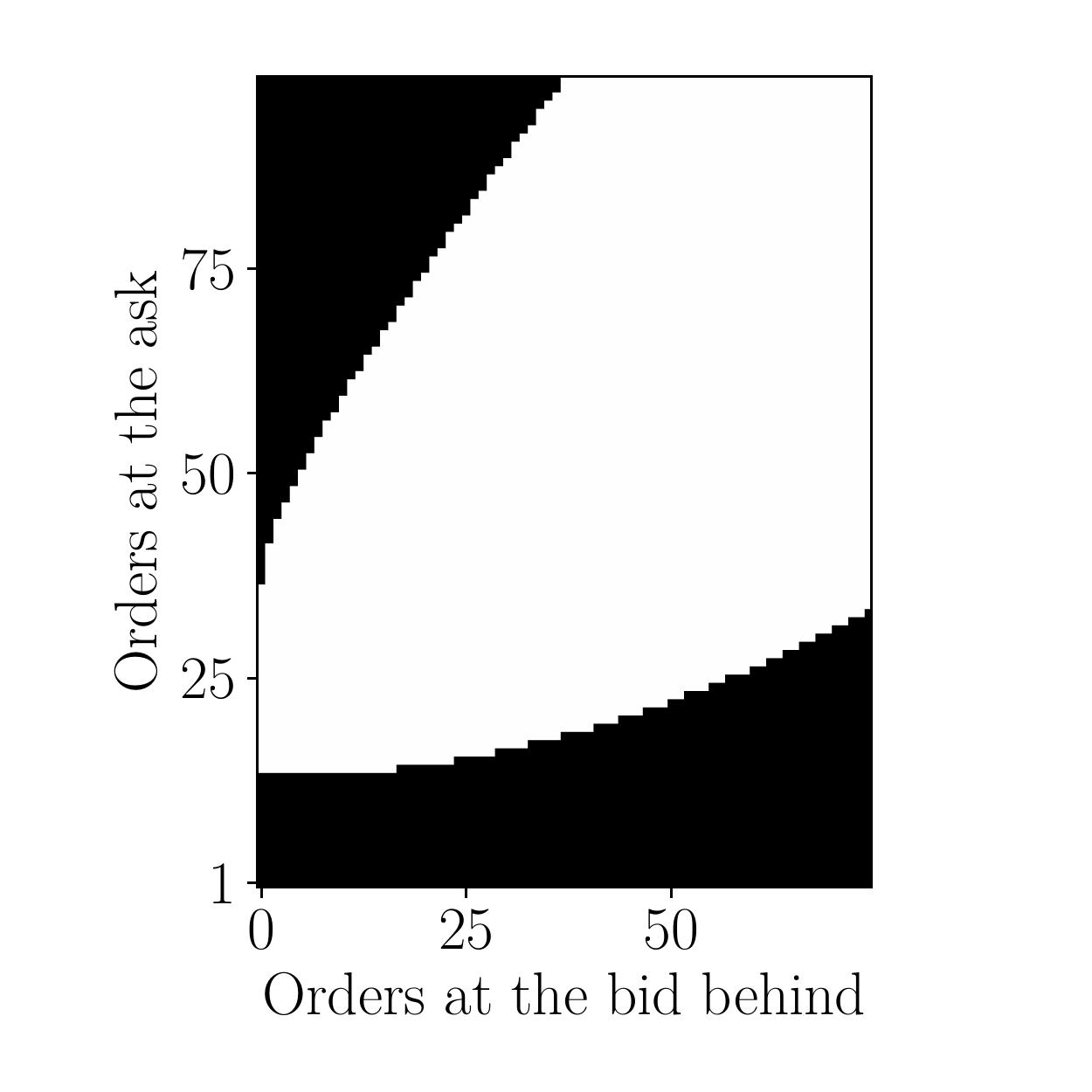} 
        \caption{Order in position 25 at the bid} 
        \label{CO_10_period_25_pos}
    \end{subfigure}
    	\hfill
    \begin{subfigure}[t]{0.45\textwidth}
        \centering
        \includegraphics[scale = 0.6]{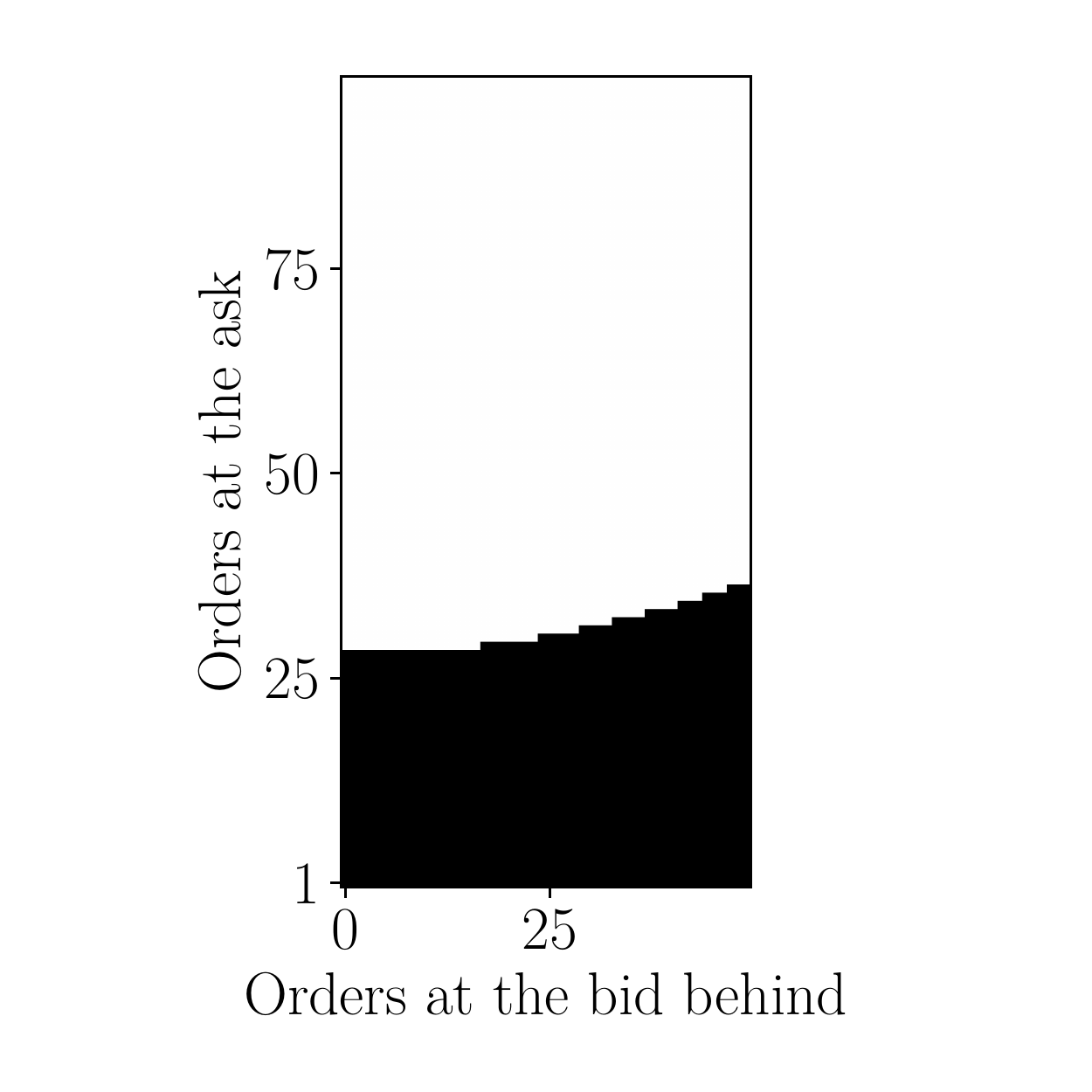} 
        \caption{Order in position 50 at the bid} 
        \label{CO_10_period_50_pos}
    \end{subfigure}
        
\captionsetup{justification=centering}
\caption{Regions of the LOB where CO submission is optimal (black) where the trader's order is in position (a) 5, (b) 10, (c) 25 and (d) 50 at the bid}
\label{co_region}        
\end{figure}

The strategy controls its LO (if any) with cancellations so that it only looks for executions when the LOB conditions are favorable. We present the cancellation regions in terms of 3 dimensions of the LOB state space (i) orders at the bid in front of our LO, (ii) orders at the bid behind our LO and (iii) orders at the ask. To better visualize it, we show in Fig. \ref{co_region} slices of the region each with a fixed number of orders at the bid in front of our LO. For simplicity we only present the case of 10 time periods remaining. The strategy looks to cancel orders when the LOB is imbalanced in any direction. It is more tolerant of positive imbalance when the order is in a good queue position. The reason for this being that if the LOB imbalance is positive, the mid-price is likely to go up, but if the order has high priority it has a good chance of being executed before the mid-price move.

Having access to all three types of orders (MO, LO, and CO) helps to mitigate non-execution and adverse selection risks in a way that is not possible if one of these order types is excluded. In the case of MOs, it is clear that without the other two order types the trader would have no way of capturing the remaining liquidity when the imbalance is significantly positive.
\subsection{Simulation}
Lehalle and Mounjid (2016) give a model for LOB dynamics and a dynamic programming
algorithm for optimizing LO placement based on imbalance. In Appendix \ref{imbalance}, we give empirical evidence that the predictive value of imbalance needs to be understood and utilized differently from how it has traditionally been understood and used.  We incorporate this new understanding into our LOB dynamics for the purposes of the simulations in this section. In addition, we incorporate more information from the LOB in the decision process and we allow a larger action space that includes market orders and
cancellations. The reason that we are able to incorporate more information and a larger action space is that we use the value iteration algorithm to approximate the optimal solution with less computational effort than direct application of dynamic programming. The second strategy used as another benchmark is derived from Jacquier and Liu (2017). They solve a similar optimal order placement problem, however their strategy uses LOs and MOs, but not cancellations. Additionally in their setup the trader is only allowed to make a decision of what type of order to place right after a mid-price change.
\par
The setup for our simulation can be summarized as follows:
\begin{itemize}
\item The objective is to buy 1 share in 10 time periods.
\item We simulate the dynamics of the LOB according to the model described in Section \ref{analysis1} for 100,000 paths.
\item The reward function is the same one described in Section \ref{placement}.
\end{itemize}
The strategies we compare in this simulation are listed below.  The first is the one proposed in this paper.  The other two are taken from the optimal execution literature. All strategies are required to submit a MO after the final mid-price move if they haven't purchased the share.
\begin{itemize}
\item All orders: the order placement strategy that uses MOs, LOs and COs.
\item No COs: this strategy only uses LOs and MOs. We also put the additional restriction of not allowing MO if a LO has been placed. Since it can't cancel stale LOs it is susceptible to adverse selection and non-execution risk.
\item No MOs: this strategy only uses LOs and COs. Since it can't aggressively submit MOs it is susceptible to non-execution risk. 
\end{itemize}
\par
The results of our simulation are shown in Table \ref{simulation}. Overall the strategy that uses all order types appears to be the best in terms of mean and standard deviation of the reward. We attribute this superior performance to more flexibility and enhanced ability to react to different LOB conditions. The strategy without COs can't react when the LOB imbalance worsens and the trader faces adverse selection risk. It also can't react to non-execution risk when the volume at the opposite side is decreasing. The strategy without MOs has similar issues. When the LO doesn't have a good queue position and the volume in the opposite side is decreasing this strategy faces non-execution risk.
\par
We also include (i) the proportion of shares purchased with LOs and MOs and (ii) the proportion of LOs that were cancelled after being submitted. As one would expect, all strategies tend to prefer buying the share with a LO. For the strategies that use COs we see that they end up cancelling most of the LOs placed. The reason for this being that the strategies are constantly adjusting their positions as market conditions become more favorable.
\begin{table}[h!]
\caption{Strategy simulation comparison}
\centering
\begin{tabular}{c | c | c | c | c | c}
Strategy & Mean reward & Std reward & Bought with LO \% & Bought with MO \% & LO cancelled \% \\
\hline
All orders & 0.35 & 0.29 & 77.62 & 22.38 & 71.34  \\
\hline
No COs & 0.21 & 0.43 & 62.16 & 37.84 & 0 \\
\hline
No MOs & 0.24 & 0.21 & 92.03 & 7.97 & 85.54 \\
\end{tabular}
\label{simulation}
\end{table}
\section{Conclusion}\label{conclusion}
We have used order flow data from the NASDAQ Historical TotalView-ITCH to study the dynamics of LOB for large tick stocks. We show that both previous orders and the relative volumes at the best quotes have a major impact in the order submission strategies of market participants. Our study shows that both effects are strong and neither should be ignored in LOB models. In particular we show that MO activity is mostly dependent on previous MOs, whereas LO and CO submission is dependent on LOB activity on the same side. Orders of all types are more common on the side favored by the volume imbalance, i.e. positive (negative) imbalance triggers buy (sell) side activity. The main contribution of our work in this area is to model these effects jointly and detail their interactions. Our approach to model the LOB dynamics differs from most of the literature since we work on discrete event time. Every point process has equivalent representations in event time and in clock time.  We choose the event time representation because it allows us to focus on those aspects of the LOB that we believe have the greatest impact on optimal trading.
\par
We conducted an empirical study to determine the time horizon for which LOB imbalance has predictive power. We show that for large-tick assets the spread follows an extremely predictable alternating cycle between 1 and 2 ticks. Imbalance of the LOB has predictive power only in the duration of its current spread cycle. When the spread equals 1, imbalance is a nontrivial predictor of which queue (bid or ask) will deplete first.  After the spread returns to size 1, the previously observed imbalance is no longer predictive of further mid-price changes. This observation does not conflict with previous empirical studies in the literature that show that imbalance is predictive of the mid-price a fixed number of milliseconds or trades in the future. Our contribution is a clear time bound (in event time) for this predictive relationship.
\par
We then propose a solution to the problem of optimal order acquisition using all order types. In our framework, the trader is looking to maximize the difference between (i) the mid-price after the immediate next bid or ask queue depletion and refill and (ii) the acquisition price. The reason for using this benchmark is our analysis (in the Appendix) that shows that the predictive power of LOB imbalance decays after a mid-price change. We derive the optimal strategy by framing the problem as a discrete time MDP. The introduction of the tools from the MDP literature is another one of our main contributions. This framework is flexible and with the appropriate changes to reward function and state space it is applicable to many other trading problems. The resulting strategy has many desirable properties, including avoiding both non-execution and adverse selection risks. It uses MOs aggressively when sell side volume is small to capture the remaining liquidity when a move up in mid-price is likely, it places LOs when the LOB is balanced and it sends COs when non-execution or adverse selection risk is high. We show, via simulation, how our strategy outperforms others that don't use all order types in terms of the expected reward. In practice, high-frequency traders usually do not have hard constraints on the types of orders they are able to place. Therefore we believe that our approach has significant applications.   
\appendix
\section{Imbalance signal}\label{imbalance}
It is known that imbalance has significant predictive power of future price changes on both clock time and event time scales.  Nevertheless, several interesting questions remain unanswered. Gould and Bonart (2016) ask if imbalance provides useful information about the direction of several price movements into the future. If so, how does this predictive power diminish over time? In this section we present an empirical analysis that provides an answer to these questions: Imbalance has strong predictive power over exactly the next 2 mid-price changes and then it decays to almost 0 for later mid-price changes. Understanding the duration of this signal is crucial for the optimization of order placement in the LOB as we described in Section \ref{placement}.  
\par 
The objective is to explain the relationship between the mid-price and imbalance. Understanding the time horizon of this relationship helps to justify the benchmark price that we used in the optimal order placement problem in Section \ref{placement}. Our claim is also compatible with an indirect assumption made on most zero-intelligence models for LOBs where the spread is assumed to be constant. In these models, after the bid or the ask queue is depleted, the mid-price is assumed to move in the direction of the queue that disappeared and the volume levels of the new best bid and ask prices are assumed to be drawn independently of the past from some given distribution. This implies that these models predict that the imbalance after a mid-price change and return of the spread to its constant size would have no power to predict any of the following mid-price changes. This is exactly the result we verify.
\subsection{Mid-price evolution}\label{midprice}
In this section, we study the evolution of the spread for large-tick stocks and show that it follows a very predictable pattern. During regular trading hours the spread has an alternating cycle between one and two ticks. Table \ref{spread} shows the transition probability of a spread of size one to a spread of size two and vice versa. From these data it is clear that the spread is almost always 2 or less, i.e. the spread rarely transitions from $1 \to 3$ or $2 \to 3$.
\begin{table}[h!]
\caption{Empirical transition probabilities of the spread}
\centering
\begin{tabular}{c | c | c }
Stock & Spread $1 \to 2$  & Spread $2 \to 1$ \\
\hline
MSFT & 99.92 \% & 99.89 \% \\
INTC  & 99.95 \% &  99.91 \%
\end{tabular}
\label{spread}
\end{table}
\par
\begin{figure}[h!]
    \begin{subfigure}[t]{0.45\textwidth}
        \centering
        \includegraphics[width=\linewidth]{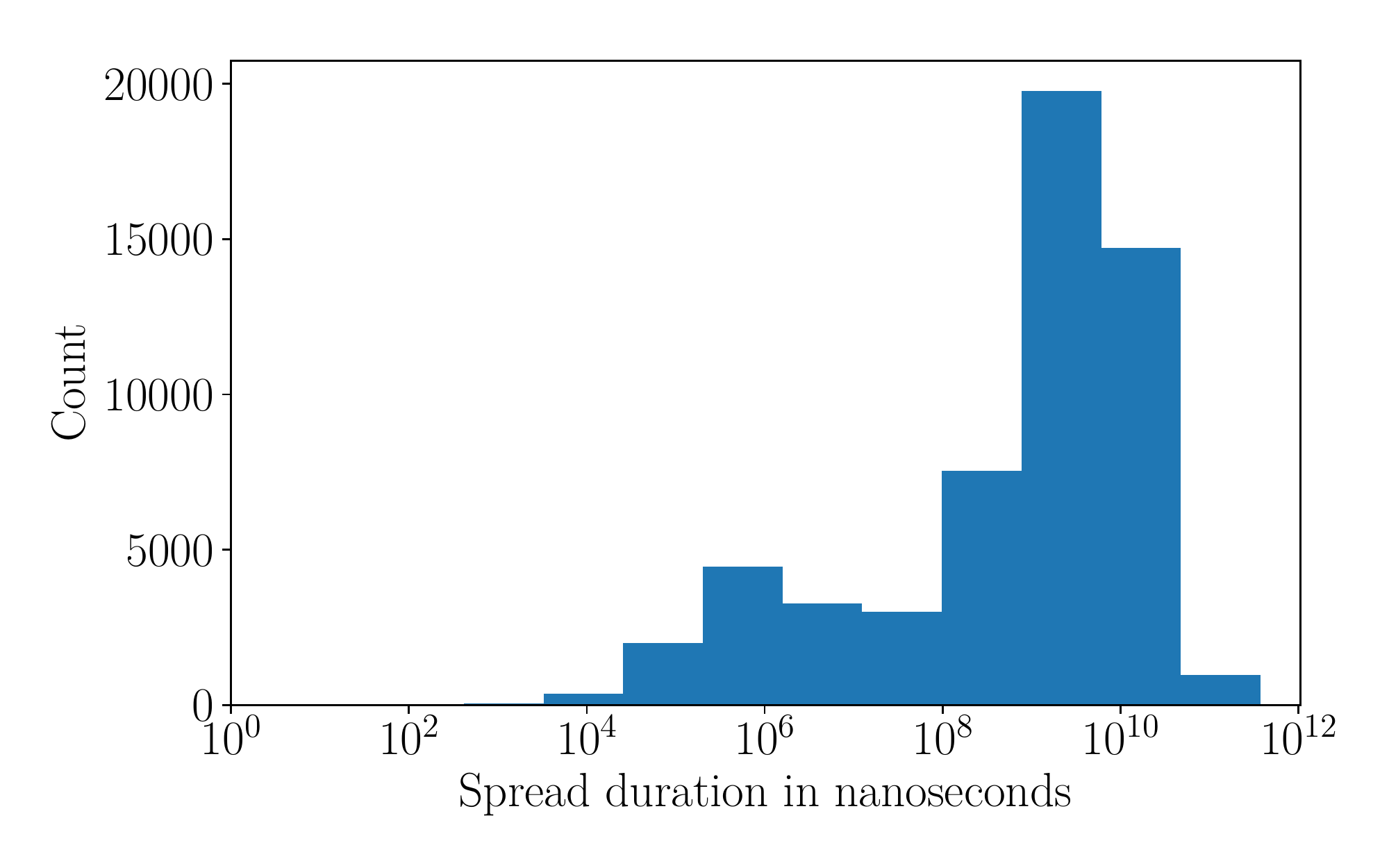} 
        \caption{Spread $=1$} 
        \label{MSFT_spread_small}
    \end{subfigure}
    	\hfill
    \begin{subfigure}[t]{0.45\textwidth}
        \centering
        \includegraphics[width=\linewidth]{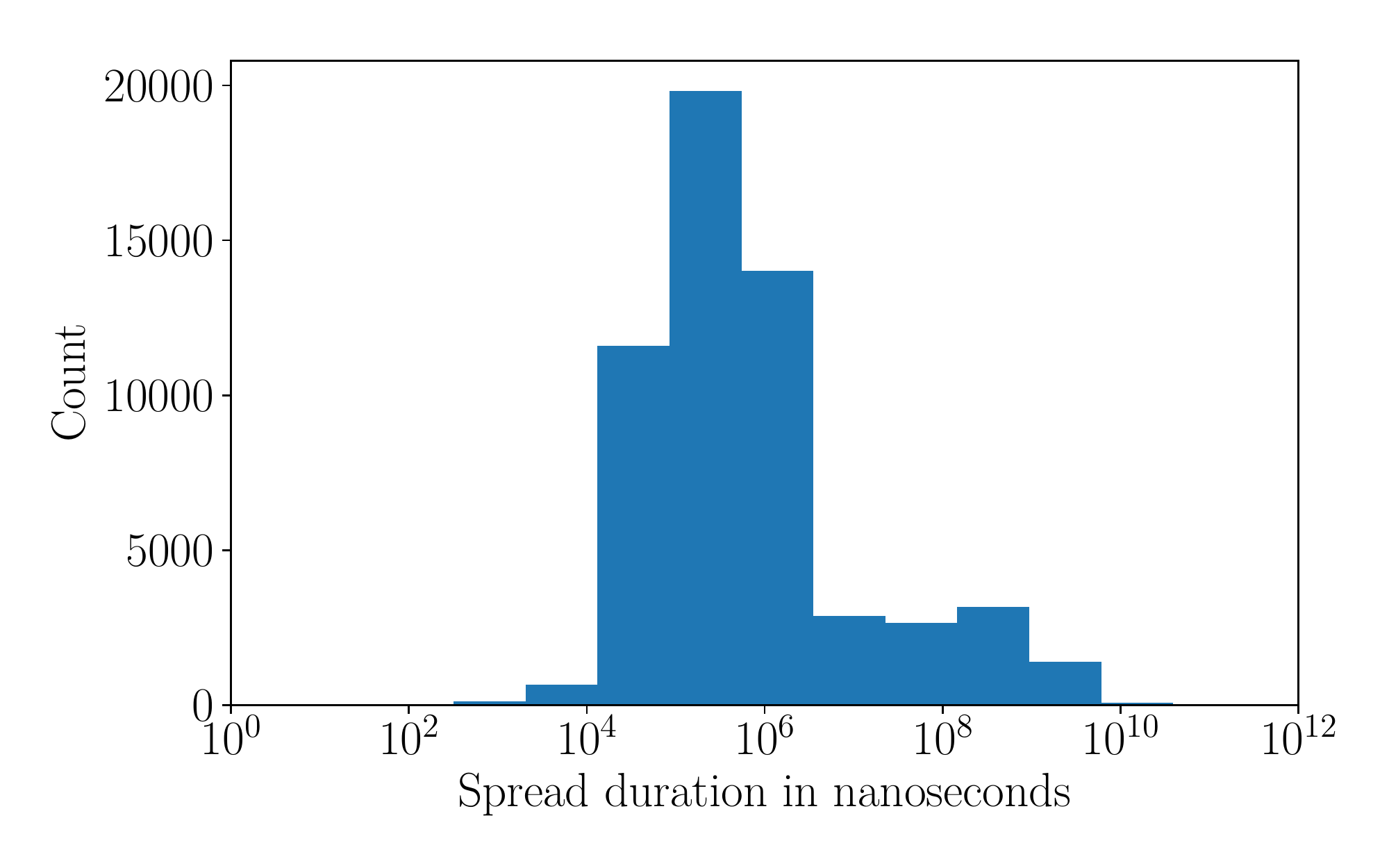} 
        \caption{Spread $>1$} 
        \label{MSFT_spread_large}
    \end{subfigure}
    \captionsetup{justification=centering}
    \caption{Histogram of the duration of the spread for MSFT stock conditional on the spread size. Spread $= 1$ (left) and spread $>1$ (right)}
    \label{MSFT_spread}
\end{figure}

 The time duration of the spread, which we measure as the time between changes in spread, is dependent on the spread value. Fig. \ref{MSFT_spread} shows the histogram of the duration of the spread for MSFT stock split into two groups, spread equal to one tick and spread greater than one tick. When the spread is equal to one tick, the duration time varies between microseconds and seconds. When the spread is greater than one tick (usually two ticks) the intensity of arrivals of limit orders between the spread is very high, closing the gap quickly. In this case the duration time of the spread rarely exceeds 1 millisecond. 
\par

\begin{figure}
\begin{subfigure}[t]{0.32\textwidth}
\begin{tikzpicture}[scale=0.65]
\draw (3.5,-1) rectangle (4.5,-5);
\draw (5.5,-3) rectangle (6.5,-5);
\draw [->](2,-5) -- ++(7.5,0);
\draw (4,-5) node[sloped]{$|$};
\draw (6,-5) node[sloped]{$|$};
\footnotesize
\draw (4,-5.5) node[below]{$ p^{b} $} ;
\draw (6,-5.5) node[below]{$ p^{a} $} ;
\draw (3.5,-4) node[left]{$V^{b}$} ;
\draw (6.5,-4) node[right]{$V^{a}$} ;
\draw (8.5,-5.5) node[right]{$Price$} ;
\end{tikzpicture}
\captionsetup{justification=centering}
\caption{Volume at the best quotes, \\ spread $=1$}
\end{subfigure}  
\hfill
\begin{subfigure}[t]{0.32\textwidth}
\begin{tikzpicture}[scale=0.65]
\draw (3.5,-1) rectangle (4.5,-5);
\draw [dotted](5.5,-3) rectangle (6.5,-5);
\draw (7.5,-2) rectangle (8.5,-5);
\draw [->](2,-5) -- ++(7.5,0);
\draw (4,-5) node[sloped]{$|$};
\draw (8,-5) node[sloped]{$|$};
\footnotesize
\draw (4,-5.5) node[below]{$ p^{b} $} ;
\draw (8,-5.5) node[below]{$ p^{a} $} ;
\draw (3.5,-4) node[left]{$V^{b}$} ;
\draw (8.5,-4) node[right]{$V^{a}$} ;
\draw (8.5,-5.5) node[right]{$Price$} ;
\end{tikzpicture}
\captionsetup{justification=centering}
\caption{Volume at the ask depleted, \\ spread $=2$}
\end{subfigure}  
\hfill
\begin{subfigure}[t]{0.32\textwidth}
\begin{tikzpicture}[scale=0.65]
\draw (5.5,-4) rectangle (6.5,-5);
\draw (7.5,-2) rectangle (8.5,-5);
\draw [->](2,-5) -- ++(7.5,0);
\draw (6,-5) node[sloped]{$|$};
\draw (8,-5) node[sloped]{$|$};
\footnotesize
\draw (6,-5.5) node[below]{$ p^{b} $} ;
\draw (8,-5.5) node[below]{$ p^{a} $} ;
\draw (5.5,-4) node[left]{$V^{b}$} ;
\draw (8.5,-4) node[right]{$V^{a}$} ;
\draw (8.5,-5.5) node[right]{$Price$} ;
\end{tikzpicture}
\captionsetup{justification=centering}
\caption{Volume at the ask replaced, \\ spread $=1$}
\end{subfigure}
\caption{Price continuation after ask being depleted}
\label{price_continuation}
\end{figure}
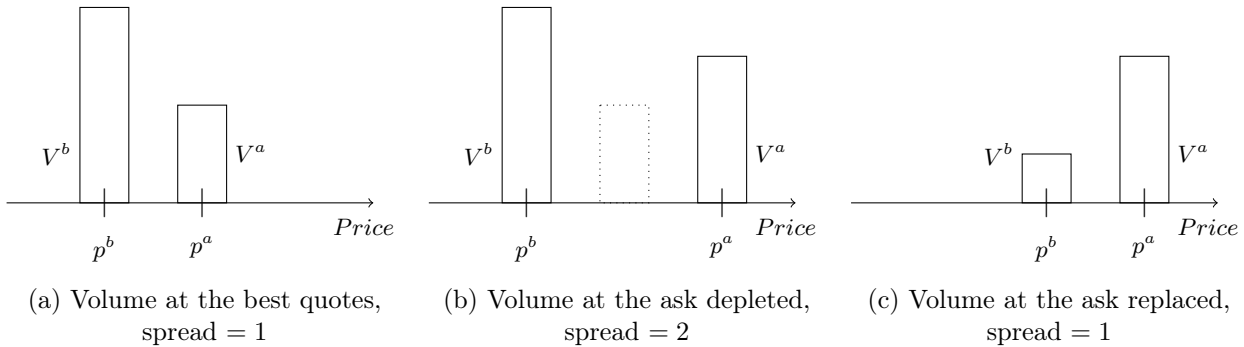

The phenomenon that will be of most interest in our study is mid-price change continuation. Mid-price continuation occurs when an emptied queue is immediately filled by orders from the opposing side. Empirically we have that the probability of this occurring is significantly larger than being refilled from orders from the same side. The diagram in Fig. \ref{price_continuation} shows the most common evolution of the mid-price and the volume at the best quotes when one of the queues is depleted.

\par
If the spread is 1 and the ask (bid) is depleted, the level of the ask is immediately filled with buy (sell) orders. In terms of mid-price changes, this can be seen as there being a higher probability of two consecutive changes in the same direction than alternating changes. In Table \ref{table_continuation} we show that mid-price continuation occurs frequently for stocks in our sample. This empirical probability drives the evolution of the mid-price in the model introduced in Section \ref{analysis1}. In this model, after a queue is depleted the mid-price goes up or down according to the empirical mid-price continuation probability. After the new mid-price is determined then the sizes of both queues are sampled independently from their empirical distributions. 

\begin{table}[h!]
\caption{Evolution of mid-price changes for large-tick stocks, starting with spread = 1}
\centering
\begin{tabular}{c | c | c | c | c | c }
Stock & $+ \to +$ & $+ \to -$ & $ - \to +$ & $ - \to -$ \\
\hline
MSFT & 81 \% & 19 \% & 21 \% & 79 \% \\
INTC  & 82 \% & 18 \% & 17 \% & 83 \% 
\end{tabular}
\label{table_continuation}
\end{table}
\subsection{Mid-price prediction}\label{logistic}
In this section we study the relationship between imbalance and future mid-price movements. In particular, we are interested in determining the horizon for which imbalance is a useful predictor. A better understanding of this horizon will allow us to better exploit this signal in the optimal order placement problem introduced in Section \ref{placement}. Let $I(t)$ denote the queue imbalance at $t$. Also denote by $t-$ and $t+$ the instants in time right before and right after time $t$ so that no LOB event occurs between $t-$ and $t$ or between $t$ and $t+$. Let $t_{1} < t_{2} < \ldots < t_{N}$ denote the times at which the mid-price changes during the period under consideration. Finally let $t_{\tau_{i}}$ be a time sampled uniformly at random between $t_{i}$ and $t_{i+1}$.
\par
The setup of our experiment can now be described as follows, we fit a nine logistic regression models, each having one of the three predictor variables $ I(t_{i-1}+), I(t_{\tau_{i}}), I(t_{i}-)$ and one of the three response variables $\Delta(t_{i+1}), \Delta(t_{i+2}), \Delta(t_{i+3})$, where $\Delta(t_{i+1}) = $ sign$(p(t_{i+1})-p(t_{i}))$. For each pair of predictor $x$ and response $y$ we model 
\begin{equation}
\hat{y} = \mathbb{P}(y = 1 | x) = \frac{1}{1 + e^{-\alpha x}}
\end{equation}
We use this probabilistic model as a classifier by taking
\begin{equation}
f(x,\alpha) = 
\begin{cases} \phantom{-}1 &\quad\text{ if } \hat{y} \geq 1/2 \\
                      -1 &\quad\text{ otherwise }
\end{cases}
\end{equation}

The results are summarized in Table \ref{logistic_table} where we show the classification accuracy of each pair of predictor and response
\begin{table}[h!]
\caption{Results of out of sample classification accuracy of logistic regression model on MSFT data}
\centering
\begin{tabular}{c | c | c | c}
& $\Delta(t_{i+1})$ & $\Delta(t_{i+2})$ & $\Delta(t_{i+3})$ \\
\hline
$I(t_{i-1}+)$ & 65 \% & 54 \% & 51\% \\
\hline
$I(t_{\tau_{i}})$ & 77 \% & 62 \% & 51\% \\
\hline
$I(t_{i}-)$ & 85 \% & 68 \% & 52\% 
\end{tabular}
\label{logistic_table}
\end{table}
\par
There are several intuitive results worth noting. Clearly the imbalance signal does have nontrivial predictive power for future price changes. Imbalance is a more accurate predictor the closer it is sampled to the mid-price change, which time is of course unknown before the fact. The most extreme case is the model $\Delta(t_{i+1}) \sim I(t_{i}-)$, where the imbalance is sampled the instant before a price change and the simple logistic regression model achieves 85\% classification accuracy. This predictive power decays in time, but more specifically, imbalance sampled at instant $t-$ seems to lose all predictive value for the third or later mid-price changes in the future. 
\par
This study gives a very well defined lifetime for the value of imbalance as a mid-price predicting signal. For $I(t_i-)$, once the mid-price has changed two times, the signal loses all predictive value for any further mid-price changes.  Nevertheless, the above result is compatible with studies that show that imbalance can predict the accumulated price change over
a fixed number of future trades as in Lehalle and Mounjid (2016) or over a fixed number milliseconds into the future as in Cartea et al. (2015c). For example, suppose that we look at a future time interval that is long enough  so that there is high probability that multiple price changes will occur. Then $I(t_i-)$ predicts $\Delta(t_{i+1})+\Delta(t_{i+2})$ very well, and the accumulated price change over the rest of the interval is equal to $\Delta(t_{i+1})+\Delta(t_{i+2})$ plus random noise, which is predicted less well as more noise accumulates.
The result is also compatible with the assumption of our LOB model that the dynamics at the best quotes follow a renewal type process after each mid-price change where the queue sizes are resampled after a mid-price change.

\end{document}